\documentclass[12pt]{article}
\usepackage{amsmath,amssymb,amsthm,amsxtra,bbm,overpic,bm,epsfig,subfigure}
\usepackage{hyperref}
\usepackage{mathrsfs}
\usepackage{enumitem}
\usepackage{graphicx}
\usepackage{color}
\usepackage{comment}
\usepackage{epstopdf}
\usepackage{float}
\usepackage{cite}
\allowdisplaybreaks[2]

\usepackage{slashed,stmaryrd,multirow}
\numberwithin{equation}{section}

\def\thefootnote{\fnsymbol{footnote}}

\usepackage{hyperref}
\usepackage{slashed,stmaryrd,orcidlink}

\usepackage{lscape}%
\usepackage{array}
\usepackage{booktabs}%

\begin{document}
	
	\vspace{0.2cm}
	
	\begin{center}
		{\Large\bf Elastic Neutrino-electron Scattering at the One-loop Level in the Standard Model}
	\end{center}
	
	\vspace{0.2cm}
	
	\begin{center}
		{\bf Jihong Huang}~{\orcidlink{0000-0002-5092-7002}}~\footnote{E-mail: huangjh@ihep.ac.cn},
		\quad
		{\bf Shun Zhou}~\orcidlink{0000-0003-4572-9666}~\footnote{E-mail: zhoush@ihep.ac.cn (corresponding author)}
		\\
		\vspace{0.2cm}
		{\small
			Institute of High Energy Physics, Chinese Academy of Sciences, Beijing 100049, China\\
			School of Physical Sciences, University of Chinese Academy of Sciences, Beijing 100049, China\\}
	\end{center}

	\vspace{0.5cm}
	
	\begin{abstract}
		In this paper, we perform a complete calculation of the differential cross section for elastic neutrino-electron scattering at the one-loop level in the Standard Model (SM), by using up-to-date values of relevant input parameters in the on-shell renormalization scheme. A careful comparison with the calculation done by Sarantakos, Sirlin and Marciano more than forty years ago in the same scheme is carried out, and an excellent agreement is found if the same values of input parameters are taken. Then we apply our results to compute the event rates for the detection of reactor antineutrinos in both JUNO and TAO experiments, which are now under construction and will soon be in operation. It should be emphasized that one-loop radiative corrections in the SM must be taken into account in the first place when searching for possible new-physics effects in the coming era of precision neutrino physics. 
	\end{abstract}
	
	\def\thefootnote{\arabic{footnote}}
	\setcounter{footnote}{0}
	
	\newpage
	\vspace{0.5cm}
	
	\section{Introduction}
	\label{sec:intro}
	
	Neutrino oscillation experiments have provided us with robust evidence that neutrinos are massive and leptonic flavor mixing is significant~\cite{ParticleDataGroup:2024cfk,Xing:2020ijf}. The origin of nonzero neutrino masses and leptonic flavor mixing definitely calls for new physics beyond the Standard Model (SM). The primary goals of next-generation neutrino oscillation experiments are to pin down the neutrino mass ordering and to discover CP violation in the leptonic sector. Furthermore, aiming to achieve these challenging measurements, future neutrino experiments will also be powerful enough to precisely measure the oscillation parameters and hopefully probe the effects of new physics that may be related to neutrino mass generation.
	
	It is obvious that a better understanding of neutrino interactions with particles in ordinary matter is beneficial for the detection of neutrinos in future experiments. Among those interactions, the elastic neutrino-electron scattering plays a special role. First, the cross sections of elastic $\nu$-$e$ scattering at the leading order (LO) and the next-to-leading order (NLO) depend solely on the electroweak (EW) part of the SM~\cite{Glashow:1961tr, Weinberg:1967tq, Salam:1968rm}. The dominant source of uncertainties in the total cross section comes from hadronic contributions involving light quarks in loops, which is about $0.2\% \sim 0.4\%$~\cite{Tomalak:2019ibg}. In contrast, the uncertainty for neutrino scattering off nucleons and nuclei, primarily due to the nucleon form factors and nuclear models, is as large as $10\% \sim 30\%$. This salient feature enables us to improve theoretical calculations with high precision. Second, there is essentially no energy threshold for elastic $\nu$-$e$ scattering to take place, so this interaction channel is particularly useful for detecting low-energy neutrinos and antineutrinos of all three flavors. In reality, however, the observation of recoiled electrons in the final states will be limited by the threshold of visible energies and relevant backgrounds in the detectors.
	
	The cross section of elastic $\nu$-$e$ scattering at the LO was first computed by 't Hooft in the EW theory in Ref.~\cite{tHooft:1971ucy}, whereas the earliest discussion on possible radiative corrections could be dated even back to 1964 in the paper by Lee and Sirlin~\cite{Lee:1964jq}. After the EW theory was proved to be renormalizable in the early 1970s~\cite{tHooft:1971akt,tHooft:1971qjg,tHooft:1972tcz,Lee:1972fj,Lee:1972ocr,Lee:1972yfa,Lee:1973fn}, several groups of authors studied one-loop radiative corrections to the $\nu$-$e$ scattering with emphases in various aspects (see, e.g., Refs.~\cite{Sirlin:1980nh,Marciano:1980pb,Sarantakos:1982bp,Aoki:1980hh,Aoki:1980ix,Aoki:1981kq,Salomonson:1974ys,Green:1980bd,Antonelli:1980zt,Bardin:1981sv} and reviews~\cite{Aoki:1982ed,Marciano:2003eq,Sirlin:2012mh}). These corrections are very important in the sense of testing the SM and also providing accurate theoretical predictions for neutrino oscillation experiments, such as the event rates of solar neutrinos in water-Cherenkov detectors~\cite{Bahcall:1995mm}.
	
	Unfortunately, when looking into previous calculations in the literature, one will immediately run into a big trouble in comparing those results, which have been obtained with different choices of renormalization schemes, input parameters and practical approximations. Such an embarrassing situation strongly motivates us to perform another complete calculation of the cross section of the elastic $\nu$-$e$ scattering at the NLO in the SM by taking account of the latest values of relevant input parameters. Two comments on the comparison between our calculation and those in the literature are in order. First, we adopt the on-shell scheme~\cite{Sirlin:1980nh,Marciano:1980pb,Sarantakos:1982bp,Aoki:1980hh,Aoki:1980ix,Aoki:1981kq} to determine the counterterms, which are introduced as usual to cancel out the ultraviolet (UV) divergences. The input parameters in our calculation include the fine-structure constant $\alpha$ in the Thomson limit, the on-shell masses $\{m^{}_W, m^{}_Z, m^{}_h, m^{}_f\}$ of weak gauge bosons $W^\pm$ and $Z$, the Higgs boson $h$ and the SM fermions $f$. The advantage of such a particular choice is that all these parameters can be extracted directly from experimental measurements. On this point, the calculation in the on-shell scheme done by Sarantakos, Sirlin and Marciano more than forty years ago in Ref.~\cite{Sarantakos:1982bp} should be compared with ours. After converting the Fermi constant $G^{}_\mu$, which was implemented in Ref.~\cite{Sarantakos:1982bp} instead of $m^{}_W$ and extracted from the precise measurement of the muon lifetime, into our input parameters, we have found an excellent agreement. Second, the most recent calculation in the $\overline{\rm MS}$ scheme in the low-energy effective theory of the SM has been accomplished by Tomalak and Hill in Ref.~\cite{Tomalak:2019ibg}. Although a detailed comparison between our results and theirs is quite nontrivial due to different theoretical frameworks, one interesting observation from comparing LO and NLO cross sections can be made. In the framework of low-energy effective theory, the NLO cross section of the $\nu$-$e$ scattering is in general smaller than the LO one, but the difference between these two is negligible in the energy range of $E_\nu^{} \lesssim 0.5~{\rm MeV}$. This can be understood by noticing that only the electromagnetic corrections arise in the effective theory, which turn out to be negative, while the one-loop weak corrections have been incorporated into the Wilson coefficients of dimension-six operators. In our calculation, the NLO cross section is larger than the LO one. However, the full NLO cross sections in both theoretical frameworks are essentially compatible with each other, as they should be. Such an observation leads us to the conclusion whether the one-loop corrections are positive or negative should be stated together with the theoretical framework of calculations. Finally, we apply our results to compute the event rates of the elastic $\nu$-$e$ scattering in the next-generation reactor neutrino experiment JUNO~\cite{JUNO:2015zny} and its near detector TAO~\cite{JUNO:2020ijm}. We find that there are approximately 10 events per day for $\overline{\nu}^{}_e$ in JUNO and nearly a thousand events per day in TAO. Both JUNO and TAO experiments will be taking data in the coming year. Although the event rates are large enough for precise measurements, the efficient reduction of relevant backgrounds will be challenging in order for both JUNO and TAO to be sensitive enough to radiative corrections.
	
	The remaining part of this paper is organized as follows. In Sec.~\ref{sec:strategy}, we specify the kinematics of elastic $\nu$-$e$ scattering in the laboratory frame and explain our strategy for the one-loop calculations. The one-loop corrections to the scattering amplitude and the analytical results for the differential cross section are presented in Sec.~\ref{sec:1loop}. Then, in Sec.~\ref{sec:num}, we give the numerical results and make a comparison with previous results in the literature. The computation of event rates for the reactor antineutrinos in JUNO and TAO is also done. Finally, we summarize our main results in Sec.~\ref{sec:sum}.

	\section{Strategy for One-loop Calculations}
	\label{sec:strategy}
	
	In this section, we explain our basic strategy for the one-loop calculations of the cross section for the elastic neutrino-electron scattering. For clarity, we take the $\nu^{}_\mu$-$e$ scattering as an explicit example, since only the neutral-current (NC) interactions are relevant. However, the generalization of the results to the cases of $\nu_e^{}$ and antineutrinos will be discussed as well. 
	
	\subsection{Kinematics and Cross Sections}
	
	For the elastic scattering $\nu^{}_\mu (k_1^{}) + e (p_1^{}) \to \nu^{}_\mu (k_2^{}) + e (p_2^{})$ under consideration, we will safely neglect neutrino masses and thus the four-momenta satisfy the on-mass-shell conditions $k_{1}^2 = k_2^2 = 0$ and $p_{1}^2 = p_2^2 = m_e^2$. Furthermore, we shall work in the laboratory frame where the initial electron is at rest and the incident neutrino is traveling along the positive direction of $z$-axis. In this case, the explicit forms of all relevant four-momenta can be written as
	\begin{eqnarray}
		k_1^{} = (E_\nu^{}, 0, 0, E_\nu^{}) \;, \quad p_1^{} = (m_e^{},0,0,0) \;, \quad
		k_2^{} = (E_\nu^{\prime},{\bf p}_\nu^{\prime}) \;, \quad p_2^{} = (E_e^{},{\bf p}_e^{}) \;,
	\end{eqnarray}
	where we further take ${\bf p}^{}_e = (|{\bf p}_e^{}|\sin\theta , 0 , |{\bf p}_e^{}| \cos\theta) $ with $\theta$ being the azimuthal angle of the final-state electron. For later convenience, we introduce the parameter $z \equiv T_e^{} / E_\nu^{} $ with $T_e^{} \equiv E_e^{} - m_e^{}$ being the kinematic energy of electron. Therefore, we arrive at the following kinematic relations
	\begin{eqnarray}
		\label{eq:four_momenta_cdot}
		k_1^{} \cdot p_1^{} &=& k_2^{} \cdot p_2^{} = m_e^{} E_\nu^{}  \;, \nonumber \\
        k_1^{} \cdot p_2^{} &=& k_2^{} \cdot p_1^{} = m_e^{} E_\nu^{} (1-z) = E_\nu^{} \left(E_e^{} - \left|{\bf p}_e^{}\right| \cos\theta\right)  \;, \\
		k_1^{} \cdot k_2^{} &=& p_1^{} \cdot p_2^{} - m_e^2 = m^{}_e E_\nu^{} z = -q^2/2  \;, \nonumber
	\end{eqnarray}
	with $q^2 \equiv (k_1^{} - k_2^{})^2 = (p_1^{} - p_2^{})^2$ being the square of momentum transfer. As long as the energy of incident neutrinos is not extremely high, e.g., $E_\nu^{} \lesssim {\cal O}({\rm PeV})$, the relation $ |q^2| \ll m_{W}^2$ or $m_Z^2$ is always fulfilled. However, we stress that the approximation $|q^2| \ll m_f^2 $ with $m_f^{}$ being the fermion mass is not always valid, especially for the light quarks. For this reason, we shall not make such an approximation in our calculations.
	
	In terms of the weak gauge coupling constant $g$ and the $W$-boson mass $m_W^{}$, the tree-level amplitude for the elastic $\nu_\mu^{}$-$e$ scattering reads
	\begin{eqnarray}
		\label{eq:M_0_mu}
		{\cal M}_0^{(\mu)} = -\frac{g^2}{4m_W^2} \left[\overline{u_{\nu_\mu^{}}^{}} (k_2^{}) \gamma^{}_\mu P_{\rm L}^{} u_{\nu_\mu^{}}^{} (k_1^{})\right] \times \left[\overline{u_e^{}} (p_2^{}) \gamma^\mu \left(c_{\rm V}^{} - c^{}_{\rm A} \gamma^{}_5 \right) u_e^{}(p_1^{})\right] \;,
	\end{eqnarray}
	where $P_{\rm L}^{} \equiv (1-\gamma^{}_5)/2$ is the left-handed projection operator, $c_{\rm V}^{} = -1/2+2\sin^2\theta^{}_{\rm w}$ and $c_{\rm A}^{} = -1/2$ refer respectively to the vector-type and axial-vector-type couplings of electrons. Here $\theta^{}_{\rm w}$ is the weak mixing angle defined by $\sin^2\theta^{}_{\rm w} \equiv 1-m_W^2/m_Z^2$. The approximation $|q^2| \ll m_W^2$ has already been made such that the terms of ${\cal O}(q^2/m_W^2)$ have been neglected. To obtain the cross section, we need to square the amplitude, average over the initial-state electron polarizations and sum over the final-state electron polarizations. After integrating over the final-state neutrino momentum, we get the differential cross section with respect to the electron recoil energy $T_e^{}$ (or equivalently $z$) with the help of Eq.~(\ref{eq:four_momenta_cdot}) as follows
	\begin{eqnarray}
		\label{eq:diff_Te_0}
		\frac{{\rm d} \sigma_0^{(\mu)}}{{\rm d} T_e^{}} = \frac{1}{E_\nu^{}} \frac{{\rm d} \sigma_0^{(\mu)}}{{\rm d} z} &=& \frac{g^4 m_e^{}}{64\pi m_W^4} \left[\frac{m_e^{} z}{E_\nu^{}} \left(c_{\rm A}^2-c_{\rm V}^2\right) + (1-z)^2 (c_{\rm A}^{} - c^{}_{\rm V})^2 + (c_{\rm A}^{} +c^{}_{\rm V})^2\right] \;.
	\end{eqnarray}
	
	On the other hand, the recoil energy of the final-state electron $T_e^{}$ can be expressed in terms of the azimuthal angle $\theta$ as 
	\begin{eqnarray}
		\label{eq:Te}
		T_e^{} = \frac{2 m_e^{} E_\nu^2 \cos^2\theta}{\left(m_e^{} + E_\nu^{}\right)^2 - E_\nu^2 \cos^2\theta} \;.
	\end{eqnarray}
	Through the kinetic relation in Eq.~(\ref{eq:Te}), we can also convert the LO and NLO differential cross sections with respect to $T^{}_e$ into those to $\cos\theta$, i.e.,
	\begin{eqnarray}
		\label{eq:diff_theta}
		\frac{{\rm d}\sigma}{{\rm d}\cos\theta} = \frac{{\rm d}\sigma}{{\rm d} z} \times \frac{\left|{\bf p}_e^{}\right|^2}{\left|{\bf p}_e^{}\right| \left(E_\nu^{} + m_e^{}\right) - E_\nu^{} E_e^{} \cos\theta} \;,
	\end{eqnarray}
	where the three-momentum of the final-state electron is $\left|{\bf p}_e^{}\right| = \sqrt{T_e^{}(T_e^{} + 2m_e^{})}$. In the laboratory frame, as the azimuthal angle is constrained to be in the range $0 \leqslant \theta \leqslant \pi/2$, we obtain
	\begin{eqnarray}
		0 \leqslant z \leqslant z_{\rm m}^{} \equiv \frac{1}{m_e^{}/(2 E_\nu^{}) + 1} \;,
	\end{eqnarray}	
	which will be used to derive the total cross section.
	
	At the one-loop level, one can always decompose the $\nu_\mu^{}$-$e$ scattering amplitude into the most general form
	\begin{eqnarray}
		\label{eq:M_1_mu}
		{\cal M}_1^{(\mu)} &=& \left[\overline{u_{\nu_\mu^{}}^{}} (k_2^{}) \gamma^{}_\mu P_{\rm L}^{} u_{\nu_\mu^{}}^{} (k_1^{})\right] \times \overline{u_e^{}} (p_2^{}) \left[\gamma^\mu \left(A - B \gamma^5\right) + C \frac{\left(p_1^{} + p_2^{}\right)^\mu}{2 m_e^{}}\right] u_e^{}(p_1^{}) \;.
	\end{eqnarray}
	The relevant coefficients $A$, $B$ and $C$ account for the radiative corrections from the gauge-boson self-energies, vertex corrections and box diagrams of ${\cal O}(g^4)$, whose explicit forms will be given in the next section. The cross section at the one-loop level contains interference terms between the LO and NLO scattering amplitudes. Implementing the result in Eq.~(\ref{eq:M_1_mu}), we can immediately obtain the most general expressions of the differential cross section
	\begin{eqnarray}
		\label{eq:diff_Te_1}
		\frac{{\rm d} \sigma_1^{(\mu)}}{{\rm d} T_e^{}} &=& \frac{g^4 m_e^{}}{64 \pi m_W^4} \left[\frac{m_e^{} z}{E_\nu^{}} \left(c_{\rm A}^2-c_{\rm V}^2\right) + (1-z)^2 (c^{}_{\rm A}-c^{}_{\rm V})^2 + (c_{\rm A}^{}+c^{}_{\rm V})^2\right]   \nonumber \\
		&& + \frac{g^2 m_e^{} }{8\pi m_W^2} \left\{ \left[z (z-2) c_{\rm A}^{} - \left(z^2-2 z+2\right) c_{\rm V}^{} +  \frac{m_e^{} z}{E_\nu^{}} c_{\rm V}^{}\right] A \right. \nonumber \\
		&& \left. + \left[z (z-2) c^{}_{\rm V} - \left(z^2-2 z+2\right) c_{\rm A}^{} -  \frac{m_e^{} z}{E_\nu^{}} c_{\rm A}^{}\right] B +  \left[2 (z-1)+\frac{m_e^{} z}{E_\nu^{}}\right] c_{\rm V}^{} C  \right\}\;.
	\end{eqnarray}
	It can be seen from Eq.~(\ref{eq:diff_Te_1}) that those terms in the first line correspond to the tree-level result of ${\cal O}(g^4)$ in Eq.~(\ref{eq:diff_Te_0}), while the radiative corrections of ${\cal O}(g^6)$ are proportional to $A$, $B$ and $C$. 
	
	For the elastic $\nu_e^{}$-$e$ scattering, one can simply make the replacements $c_{\rm V}^{} \to c_{\rm V}^{} + 1$ and $c_{\rm A}^{} \to c_{\rm A}^{} + 1$ in Eqs.~(\ref{eq:diff_Te_0}) and (\ref{eq:diff_Te_1}) since there are extra contributions via the charged-current (CC) interactions. In addition, at the one-loop order, three coefficients $A$, $B$ and $C$ will receive the corrections also from the CC interactions, which need to be taken into account. In the case of antineutrinos, one could make the changes $c_{\rm A}^{} \to - c_{\rm A}^{}$ and $B \to -B$ to the expressions for the neutrino cross section in Eqs.~(\ref{eq:diff_Te_0}) and (\ref{eq:diff_Te_1}).
	
	\subsection{On-shell Renormalization}
	
	The one-loop renormalization of the SM in the on-shell scheme has been presented in many excellent monographs and review articles~\cite{Aoki:1982ed,Bohm:1986rj,Hollik:1988ii,Denner:1991kt,Bohm:2001yx,Sirlin:2012mh,Denner:2019vbn}, and a recent summary can be found in the Appendix of Ref.~\cite{Huang:2023nqf}. In this work, we closely follow the notations and conventions in Ref.~\cite{Huang:2023nqf}, and some basics are briefly mentioned in this subsection. 
	
	After introducing the gauge-fixing terms and the Faddeev-Popov ghosts into the Lagrangian, one can derive the Feynman rules of the SM, for which the ’t Hooft-Feynman gauge is chosen for simplicity. The bare fields and parameters are decomposed into the renormalized ones and the corresponding counterterms. Then, the Higgs tadpole diagrams, the self-energies of gauge-bosons and fermions, and the one-loop vertex functions can be directly calculated, where we will inevitably encounter the UV divergence. The dimensional regularization, where the space-time dimension is set to $d = 4-2\epsilon$, is adopted to separate the divergent terms from finite ones. The UV divergence will be denoted as $\Delta \equiv 1/\epsilon - \gamma^{}_{\rm E} + \ln(4\pi)$, with $\gamma^{}_{\rm E} \approx 0.577$ being the Euler-Mascheroni constant. Once the counterterms are determined by the on-shell renormalization conditions and a particular set of renormalized parameters are chosen as input, we will be able to compute the UV-finite $S$-matrix elements for physical processes of our interest. We choose the input parameters as the fine-structure constant $\alpha$, the $W$-boson mass $m_W^{}$, the $Z$-boson mass $m_Z^{}$, the Higgs-boson mass $m_h^{}$, and the charged-fermion masses $m_f^{}$. The weak mixing angle is defined as $\cos\theta_{\rm w}^{} \equiv m_W^{}/m_Z^{}$, and the abbreviations $c \equiv \cos\theta_{\rm w}^{}$ and $s \equiv \sin\theta_{\rm w}^{}$ will be frequently used in this work. The electromagnetic coupling constant $ e = \sqrt{4\pi\alpha}$ is related to weak gauge coupling $g$ via $e = g s$.
	
	The contributions to the elastic $\nu$-$e$ scattering amplitudes at the one-loop level contain three parts, i.e., the self-energy corrections of the gauge bosons, the vertex corrections and the box diagrams. All the UV-divergent terms arising from one-loop diagrams will be canceled by the counterterms listed in the Appendix of Ref.~\cite{Huang:2023nqf}. The one-loop amplitudes are calculated with the help of the publicly available package {\tt Package-X}~\cite{Patel:2015tea,Patel:2016fam}, and expressed in terms of the Passarino-Veltman functions~\cite{Passarino:1978jh}. We shall introduce the notation $x_i^{} \equiv m_i^2/m_W^2$ with ``$i$" referring to the particle type to simplify the analytical expressions. Since the electron mass is much smaller than the weak gauge-boson masses, we can safely neglect the terms of ${\cal O}(x_e^{})$. Besides, the flavor mixing of quarks is ignored as the Cabibbo-Kobayashi-Maskawa (CKM) matrix~\cite{Cabibbo:1963yz,Kobayashi:1973fv} is approximately diagonal and the CC vertices involving a pair of quarks that not in the same isospin-doublet are significantly suppressed.
	
	\begin{figure}[t]
		\centering
		\includegraphics[scale=1]{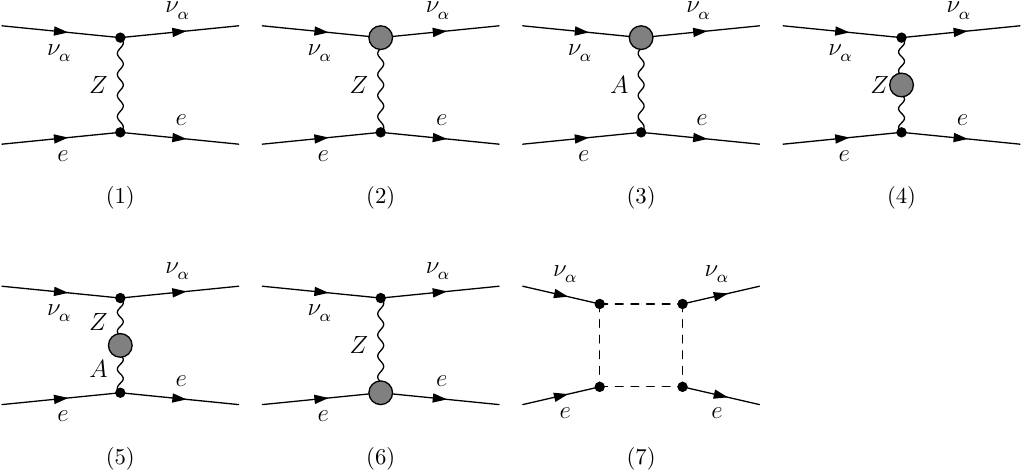}
		\caption{The relevant Feynman diagrams of the elastic $\nu_\alpha^{}$-$e$ scattering via the NC process at the tree level (1) and the one-loop level (2)-(7). The shaded circle represents the radiative corrections to vertices and corresponding propagators. The dashed box in (7) indicates all possible realizations of internal lines.}
		\label{fig:NC}
	\end{figure}
		
	The relevant diagrams of the $\nu_\alpha^{}$-$e$ scattering via the NC process are plotted in Fig.~\ref{fig:NC}. For $\nu_{\mu,\tau}^{}$, Fig.~\ref{fig:NC} describes all the one-loop corrections since they only participate in the NC interactions. It is clear that the one-loop corrections can be divided into the following six parts, i.e., self-energies of the $Z$-boson and the $A$-$Z$ mixing, the corrections to the $\nu^{}_\alpha$-$\nu^{}_\alpha$-$Z$, $\nu^{}_\alpha$-$\nu^{}_\alpha$-$A$ and $e$-$e$-$Z$ vertices, and NC box diagrams. Therefore, the coefficients $A$ in Eq.~(\ref{eq:M_1_mu}) can be written more explicitly as
	\begin{eqnarray}
		\label{eq:A-NC}
		A_{\rm NC}^{} = A_{Z}^{} + A_{AZ}^{} + A_{\nu^{}_\alpha \nu^{}_\alpha Z}^{} + A_{\nu^{}_\alpha \nu^{}_\alpha A}^{} + A_{eeZ}^{} + A_{\rm \Box,NC}^{} \;.
	\end{eqnarray}
	Similar expressions can also be obtained for the other coefficients $B$ and $C$. 
	
	\begin{figure}[t]
		\centering
		\includegraphics[scale=1]{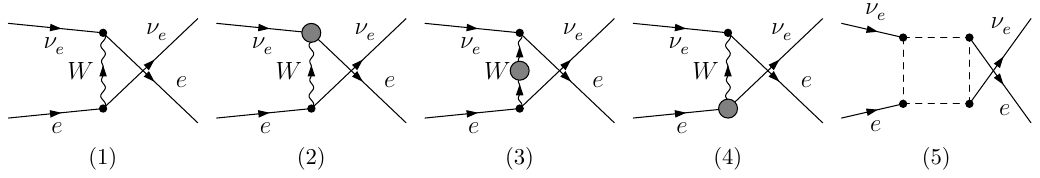}
		\caption{The relevant Feynman diagrams of the $\nu_e^{}$-$e$ scattering via the CC process at the tree level (1) and the one-loop level (2)-(5). The notations are the same as those in Fig.~\ref{fig:NC}.}
		\label{fig:CC}
	\end{figure}
	
	For electron neutrinos, the CC process also provides three extra contributions to $A$ from the $W$-boson self-energy, the $\nu_e^{}$-$e$-$W$ vertex corrections and the CC box diagrams, which reads
	\begin{eqnarray}
		\label{eq:A-CC}
		A_{\rm CC}^{} = A_{\rm NC}^{} + A_{W}^{} + 2 \times A_{\nu^{}_e e W}^{} + A_{\rm \Box,CC}^{} \;.
	\end{eqnarray}
	The factor of two in front of $A_{\nu^{}_e e W}^{}$ comes from the fact that there are two identical vertex corrections that need to be taken into account. In Fig.~\ref{fig:CC}, we plot the relevant Feynman diagrams at the tree level (1) and the one-loop level (2)-(5). 
	
	Since our calculations are performed at the cross-section level, the bremsstrahlung of photons from the initial- or final-state electron should also be taken into account in order to get rid of the infrared (IR) divergence and mass singularities~\cite{Kinoshita:1962ur,Lee:1964is}. In this work, the real emission of both soft and hard photons are included, which not only cancels out the IR divergence from the massless limit of the photon $\lambda\to 0$ but also eliminates the collinear divergence caused by $m_e^{}\to 0$. The technique to include hard-photon bremsstrahlung has been discussed in detail in Refs.~\cite{Ram:1967zza,Passera:2000ug}. We adopt the approach in Ref.~\cite{Sarantakos:1982bp} and improve it so that its applicability is not limited to the extreme relativistic regime of the final-state electron.
	
	The contributions from self-energies, vertices and box diagrams on the right-hand side of Eqs.~(\ref{eq:A-NC}) and (\ref{eq:A-CC}) will be given in Sec.~\ref{sec:1loop}, where the cross section for the processes with real photon emission are also calculated. The total cross sections at the one-loop level will be evaluated with the latest values of input parameters in Sec.~\ref{sec:num}. Given the comprehensive introduction to the renormalization of the SM in the Appendix of Ref.~\cite{Huang:2023nqf}, including the definition of counterterms, the on-shell renormalization conditions, the analytical expressions of the self-energies and vertex counterterms, and most of one-loop Feynman diagrams, we will not repeat them in this paper. Instead, we refer to the corresponding formulae or figures when needed, except for those not given therein. To distinguish the numberings of equations and figures in Ref.~\cite{Huang:2023nqf} from those in this paper, we {\it underline} all the equations and figures quoted from Ref.~\cite{Huang:2023nqf}.

	\section{Analytical Results}
	\label{sec:1loop}
	
	In this section, we provide the analytical expressions of relevant coefficients that are required to calculate the NLO cross section. The specific forms of the self energies and counterterms involved in these expressions can be found in Ref.~\cite{Huang:2023nqf}. In addition, we have compared and verified the consistency of these expressions for each part with the results given in Refs.~\cite{Marciano:1980pb,Aoki:1982ed,Sarantakos:1982bp}.
	
	\subsection{Self-energy Contributions}
	
	Let us first examine the self-energy contributions. For the NC process, the radiative corrections are illustrated in Fig.~\ref{fig:NC}-(4) and (5), where the shaded circle denotes all one-loop contributions. In the limit of $|q^2| \ll m_Z^2$, the renormalized self-energy of $Z$-boson $\widehat{\Sigma}_{\rm T}^{Z} (q^2)$ reads
	\begin{eqnarray}
		\widehat{\Sigma}_{\rm T}^{Z} (q^2) = \Sigma_{\rm T}^{Z} (q^2) +\delta m_Z^2 + m_Z^2 \delta Z_{Z}^{} \;, 
	\end{eqnarray}
	where the counterterms and the specific form of the $Z$-boson self-energy $\Sigma_{\rm T}^{Z} (q^2)$ can be determined from Eqs.~\underline{(A1)}, \underline{(A2)}, \underline{(A4)} and \underline{(A11)}, respectively. The one-loop diagrams for the $Z$-boson self-energy and the corresponding counterterm are plotted in Fig.~\underline{4}. Therefore, the contributions to the coefficients in Eq.(\ref{eq:M_1_mu}) are given by
	\begin{eqnarray}
		\frac{A^{}_Z}{c_{\rm V}^{}} = \frac{B^{}_Z}{c_{\rm A}^{}} = \frac{g^2}{4 c^2 m_Z^4}  \widehat{\Sigma}_{\rm T}^Z(q^2) \;, \quad C_Z^{} = 0 \;.
	\end{eqnarray}
	
	Similarly, the renormalized $A$-$Z$ mixing self-energy $\widehat{\Sigma}_{\rm T}^{AZ} (q^2)$ can be expressed as
	\begin{eqnarray}
		\label{eq:AZ_renormal}
		\widehat{\Sigma}_{\rm T}^{AZ} (q^2) = \Sigma_{\rm T}^{AZ} (q^2) -\frac{1}{2}\left[\left(\delta Z_{ZA}^{} + \delta Z_{AZ}^{}\right) q^2 - m_Z^2 \delta Z_{ZA}^{} \right] \;.
	\end{eqnarray}
	The wave-function counterterms $\delta Z_{ZA}^{}$ and $\delta Z_{AZ}^{}$ are defined in Eqs.~\underline{(A2)} and \underline{(A5)}, while $\Sigma_{\rm T}^{AZ} (q^2)$ is expressed in Eq.~\underline{(A14)}, and the one-loop diagrams are plotted in the Fig.~\underline{6}. Since the $e$-$e$-$A$ vertex only contains the vector-current interaction, the contributions to the coefficients are
	\begin{eqnarray}
		A_{AZ}^{} = \frac{g^2 s}{2 c q^2 m_Z^2} \widehat{\Sigma}_{\rm T}^{AZ} (q^2) \;, \quad B_{AZ}^{} = C_{AZ}^{} = 0 \;.
	\end{eqnarray}
	We notice that the photon propagator ${\rm i}/q^2$ contains a pole at $q^2 = 0$, so all the terms proportional to $q^2$ in the square brackets of Eq.~(\ref{eq:AZ_renormal}) should not be ignored. These terms ensure that such a pole is canceled out by the renormalized self-energy $\widehat{\Sigma}_{\rm T}^{AZ} (q^2)$.
	
	For the CC interaction, the $W$-boson self-energy in Fig.~\ref{fig:CC}-(3) contributes to the corresponding coefficients as
	\begin{eqnarray}
		A_W^{} = B_W^{} = \frac{g^2}{4 m_W^4} \widehat{\Sigma}_{\rm T}^W(q^2) \;, \quad C_W^{} = 0 \;,
	\end{eqnarray} 
	where the renormalized self-energy is defined as usual
	\begin{eqnarray}
		\widehat{\Sigma}_{\rm T}^{W} (q^2) = \Sigma_{\rm T}^{W} (q^2) +\delta m_W^2 + m_W^2 \delta Z_{W}^{} \;, 
	\end{eqnarray}
	with the counterterms in Eqs.~\underline{(A1)}, \underline{(A2)} and \underline{(A4)} and $\Sigma_{\rm T}^{W} (q^2)$ from Eq.~\underline{(A12)}. The one-loop self-energy of the $W$-boson is plotted in Fig.~\underline{5}.

	\subsection{Vertex Contributions}
	
	\subsubsection{$\nu_\alpha^{}$-$\nu_\alpha^{}$-$Z$ and $\nu_\alpha^{}$-$\nu_\alpha^{}$-$A$ Vertices}
	
	The Feynman diagrams of vertex corrections in the NC part have been depicted in Fig.~\ref{fig:NC}-(2), (3) and (6). First, we focus on the radiative corrections to the $\nu_\alpha^{}$-$\nu_\alpha^{}$-$Z$ vertex (see Fig.~\underline{8} with $f\to\nu_\alpha^{}$ and discussions therein). They contribute to the coefficients as 
	\begin{eqnarray}
		\frac{A_{\nu^{}_\alpha \nu^{}_\alpha Z}^{}}{c_{\rm V}^{}} &=& \frac{B_{\nu^{}_\alpha \nu^{}_\alpha Z}^{}}{c_{\rm A}^{}} \nonumber \\
		&=& - \frac{1}{(4\pi)^2 }\frac{g^4}{16 c^2 m_W^2}\left[ \left(8 c^4+2 c^2+1\right)\Delta +2c^2 \left(4 c^2+1\right) \ln \left(\frac{\mu ^2}{m_W^2}\right) \right. \nonumber \\
		&& \left. + \ln \left(\frac{\mu ^2}{m_Z^2}\right) -\frac{1}{2}- c^2\right] - \frac{1}{(4\pi)^2}\frac{g^4  }{ 16 m_W^2} x_\alpha^{} \left[\Delta + \ln  \left(\frac{\mu ^2}{m_W^2}\right) -4 \ln x_\alpha^{} -\frac{27}{2}\right]  \nonumber \\
		&&  -\frac{g^2 s }{2m_Z^2 c} g_{\nu_\alpha^{}}^-  \left(\frac{\delta g_{\nu_\alpha^{}}^-}{g_{\nu_\alpha^{}}^-} + \frac{1}{2} \delta Z_{Z}^{} + \delta Z_{\rm \nu_\alpha^{}}^{\rm L}\right) \;, \nonumber \\
		C_{\nu^{}_\alpha \nu^{}_\alpha Z}^{} &=& 0 \;.
	\end{eqnarray}
	The terms in the last parentheses come from the corresponding counterterm of $\nu_\alpha^{}$-$\nu_\alpha^{}$-$Z$ vertex, where the definitions of the fermion wave-function counterterms $\delta Z_{\rm \nu_\alpha^{}}^{\rm L}$, coefficients $g_{\nu_\alpha^{}}^-$ and $\delta g_{\nu_\alpha^{}}^-$ can be read off accordingly from Eqs.~\underline{(A6)} and \underline{(A22)} with $f\to \nu^{}_\alpha$, and the fermion self-energy is expressed in Eq.~\underline{(A15)}. In addition, to derive the explicit expressions of $\delta g_{\nu_\alpha^{}}^-$, one may need the renormalization constant of the electric charge $\delta Z_e^{}$ and the counterterm of the weak mixing angle $\delta s/s$ from Eqs.~\underline{(A8)} and \underline{(A9)}, respectively. In order to cancel out the UV-divergent part dependent on the neutrino flavor in the one-loop amplitudes, the flavor-dependent terms of $\delta Z_{\rm \nu_\alpha^{}}^{\rm L}$ should be kept up to ${\cal O}(x_\alpha^{})$.
	
	\begin{figure}[t]
		\centering
		\includegraphics[scale=1]{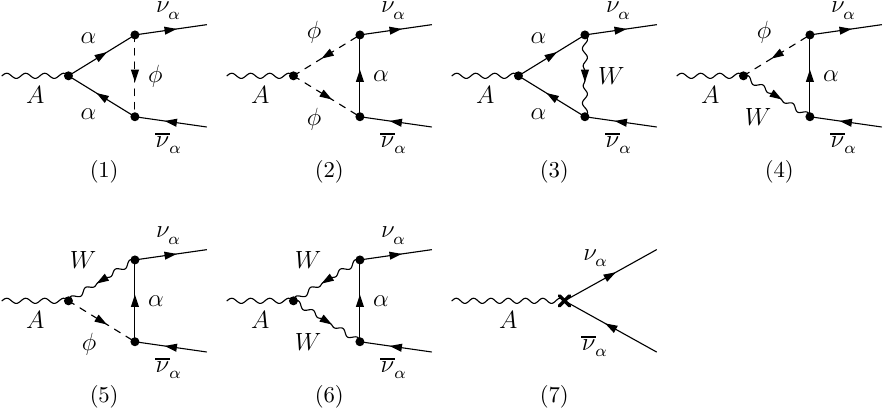}
		\caption{The one-loop diagrams of the $\nu^{}_\alpha$-$\nu_\alpha^{}$-$A$ vertex and the corresponding counterterm in the diagram (7).}
		\label{fig:nu_nu_A}
	\end{figure}
	
	The one-loop diagrams of the $\nu^{}_\alpha$-$\nu_\alpha^{}$-$A$ vertex are plotted in Fig.~\ref{fig:nu_nu_A} with the corresponding counterterms. In such corrections, the flavor-dependent terms are not only proportional to $x_\alpha^{}$ or $x_\alpha^{} \ln x_\alpha^{}$ but also to $\ln x_\alpha^{}$, which contributes the most to the flavor-dependent corrections. Similar to the case of the $A$-$Z$ mixing, there is only the vector-current interaction and the photon propagator contributes a pole as ${\rm i}/q^2$. Therefore, those terms proportional to the momentum-transfer square should be maintained. After including the counterterms, at the order of ${\cal O}(x_\alpha^{})$, the renormalized vertex corrections are proportional to $q^2$ and the total amplitude does not contain such a pole anymore. The contributions to the coefficients read
	\begin{eqnarray}
		A_{\nu_\alpha^{} \nu_\alpha^{} A}^{} &=& - \frac{1}{(4\pi)^2} \frac{g^4 s^2}{m_W^2} \left\{ \frac{m_W^2}{q^2} \left[\Delta + \ln \left( \frac{\mu^2}{m_W^2} \right) \right] - \frac{1}{3} + 2 \int_{0}^{1} {\rm d}x\ x(1-x) \ln\left[\frac{m_\alpha^2 - q^2 x (1-x)}{m_W^2}\right]\right\} \nonumber \\
		&& + \frac{1}{(4\pi)^2} \frac{g^4 s^2 }{m_W^2} x_\alpha^{} \left\{\frac{29}{36} + \int_{0}^{1} {\rm d}x\ (1-2x+2x^2)\ln\left[\frac{m_\alpha^2 - q^2 x (1-x)}{m_W^2}\right]\right\} \nonumber \\
		&& -\frac{ g^2 s^2}{2 q^2} g_{\nu_\alpha^{}}^{-} \delta Z_{ZA}^{}   \;, \nonumber \\
		B_{\nu_\alpha^{} \nu_\alpha^{} A}^{} &=& C_{\nu_\alpha^{} \nu_\alpha^{} A}^{} =0 \;,
	\end{eqnarray}
	where the counterterm in the last line cancels the pole at $q^2 = 0$ in the first line. As previously noted, the approximation $|q^2| \ll m_f^2$ is not universally valid for all fermions. Therefore, we retain them in the integral and perform numerical calculations of the total cross section.
	
	\subsubsection{$e$-$e$-$Z$ Vertex}
	
	The one-loop contributions to the $e$-$e$-$Z$ vertex corrections are plotted in Fig.~\underline{8} with $f \to e$. They contain two parts: the corrections with exchanging a virtual photon between two electrons in Fig.~\underline{8}-(7) and the other corrections with exchanging weak gauge bosons. In the early literature, the former is usually called the QED correction and the latter the EW correction. Therefore, we can also divide the coefficients into three parts as $A_{eeZ}^{} = A_{eeZ}^{\rm QED} + A_{eeZ}^{\rm EW} + A_{eeZ}^{\rm C}$ and $B_{eeZ}^{} = B_{eeZ}^{\rm QED} + B_{eeZ}^{\rm EW} + B_{eeZ}^{\rm C}$, where the last part stands for the corresponding counterterm, while $C_{eeZ}^{}$ stems only from the QED correction and is UV-finite. More explicitly, we have the EW corrections
	\begin{eqnarray} 
		\label{eq:eeZEW}
		A_{eeZ}^{\rm EW} &=& \frac{1}{(4\pi)^2} \frac{g^4 }{32 c^2 m_W^2} \left[ \left(16 c^6-24 c^4+28 c^2-9\right) \Delta + 2 \left(6 c^2-1\right) c^2 \ln \left(\frac{\mu ^2}{m_W^2}\right) \right. \nonumber \\
		&& \left. + \left(16 c^6-36 c^4+30 c^2-9\right) \ln \left(\frac{\mu ^2}{m_Z^2}\right) -8 c^6+16 c^4-14 c^2+\frac{9}{2} \right] \;, \nonumber \\
		B_{eeZ}^{\rm EW} &=& \frac{1}{(4\pi)^2} \frac{g^4}{32 c^2 m_W^2} \left[\left(24 c^4-20 c^2+7\right) \Delta +  \left(12 c^4-18 c^2+7\right) \ln \left(\frac{\mu ^2}{m_Z^2}\right) \right. \nonumber \\
		&& \left. +2c^2 \left(6 c^2- 1 \right) \ln \left(\frac{\mu ^2}{m_W^2}\right) -8 c^4+10 c^2-\frac{7}{2} \right]  \;,
	\end{eqnarray}
	and the QED corrections
	\begin{eqnarray}
		A_{eeZ}^{\rm QED} &=& - \frac{1}{(4\pi)^2} \frac{g^4 }{8 m_W^2} \left(4 c^4-7 c^2 + 3 \right) {\cal F}  \;, \nonumber \\
		B_{eeZ}^{\rm QED} &=& \frac{1}{(4\pi)^2} \frac{g^4 s^2}{8  m_W^2} {\cal G}  \;, \nonumber \\
		C_{eeZ}^{} &=& \frac{1}{(4\pi)^2} \frac{g^4}{4 m_W^2} \frac{ m_e^{}}{\left|{\bf p}_e^{}\right|} \left(4 c^4-7 c^2+3\right)  \ln \left(\frac{E_e^{} + |{\bf p}_e^{}| }{m_e^{}}\right)  \;,
	\end{eqnarray}
	where we define
	\begin{eqnarray}
		{\cal F} &=& \Delta + \ln\left(\frac{\mu^2}{m_e^2}\right) + \frac{3 \left|{\bf p}_e^{}\right|}{T_e^{}} \ln \left(\frac{E_e^{} + |{\bf p}_e^{}| }{m_e^{}}\right) \nonumber \\
		&& - \frac{2 E_e^{}}{|{\bf p}_e^{}|} \left\{\frac{1}{2} \ln \left(\frac{E_e^{} + |{\bf p}_e^{}| }{m_e^{}}\right) \ln \left[\frac{2 \left(E_e^{} + m_e^{} \right) }{m_e^{}}\right]-\ln \left(\frac{E_e^{} + |{\bf p}_e^{}| }{m_e^{}}\right) \ln \left(\frac{\lambda ^2}{m_e^2}\right)-\Phi \right\} \;, \nonumber \\
		{\cal G} &=& \Delta + \ln \left(\frac{\mu ^2}{m_e^2}\right) +\frac{3E_e^{} - m_e^{}}{\left|{\bf p}_e^{}\right|} \ln \left(\frac{E_e^{} + |{\bf p}_e^{}| }{m_e^{}}\right) \nonumber \\
		&& - \frac{2 E_e^{}}{|{\bf p}_e^{}|} \left\{\frac{1}{2} \ln \left(\frac{E_e^{} + |{\bf p}_e^{}| }{m_e^{}}\right) \ln \left[\frac{2 \left(E_e^{} + m_e^{} \right) }{m_e^{}}\right]-\ln \left(\frac{E_e^{} + |{\bf p}_e^{}| }{m_e^{}}\right) \ln \left(\frac{\lambda ^2}{m_e^2}\right)-\Phi \right\} \;.
	\end{eqnarray}
	The function $\Phi$ can be expressed with the help of the Spence function\footnote{In the literature, the dilogarithm function is sometimes denoted as ${\rm Li}_2^{}(x)$. Notice that the definition of the Spence function in Eq.~(\ref{eq:spence}) differs from that in Refs.~\cite{Sarantakos:1982bp,Marciano:2003eq} by a minus sign.}
	\begin{eqnarray}
		\label{eq:spence}
		{\rm Sp}(x) \equiv -\int_{0}^{x} \frac{\ln(1-t)}{t} \  {\rm d}t\; ,
	\end{eqnarray}
	as 
	\begin{eqnarray}
		\label{eq:Phi}
		\Phi = {\rm Sp}\left(\frac{E_e^{} - m_e^{} + \left|{\bf p}_e^{}\right|}{2 \left|{\bf p}_e^{}\right|}\right) - {\rm Sp}\left(\frac{-E_e^{} + m_e^{} + \left|{\bf p}_e^{}\right|}{2 \left|{\bf p}_e^{}\right|}\right) \;.
	\end{eqnarray}
	Finally, the corresponding counterterms are 
	\begin{eqnarray}
		A_{eeZ}^{\rm C} &=& -\frac{g^2 s}{4m_Z^2 c} \left[g_e^- \left(\frac{\delta g_e^-}{g_e^-} + \frac{1}{2} \delta Z_{Z}^{} + \delta Z^{\rm L}_{e}\right) + g_e^+ \left(\frac{\delta g_e^+}{g_e^+} + \frac{1}{2} \delta Z_{Z}^{} + \delta Z^{\rm R}_{e}\right) - \delta Z_{AZ}^{}\right] \;, \nonumber \\
		B_{eeZ}^{\rm C} &=& -\frac{g^2 s}{4m_Z^2 c} \left[g_e^- \left(\frac{\delta g_e^-}{g_e^-} + \frac{1}{2} \delta Z_{Z}^{} + \delta Z^{\rm L}_{e}\right) - g_e^+ \left(\frac{\delta g_e^+}{g_e^+} + \frac{1}{2} \delta Z_{Z}^{} + \delta Z^{\rm R}_{e}\right) \right] \;.
	\end{eqnarray}
	The definitions of the counterterms $\delta g_e^\pm$ together with the coefficients $g_e^\pm$ can be found from Eqs.~\underline{(A21)} and \underline{(A22)}. The electron wave-function counterterm $\delta Z_{e}^{\rm L,R}$ is defined in Eq.~\underline{(A6)} with $f\to e$ together with the self-energy in Eq.~\underline{(A15)}.

	\subsubsection{$\nu_e^{}$-$e$-$W$ Vertex}
	
	As illustrated in Fig.~\ref{fig:CC}-(2) and (4), the $\nu_e^{}$-$e$-$W$ vertex appears in the CC interaction for elastic $\nu_e^{}$-$e$ scattering. The relevant one-loop diagrams are plotted in Fig.~\underline{9}, and the contributions to coefficients read 
	\begin{eqnarray}
		A_{\nu^{}_e e W}^{} &=& B_{\nu^{}_e e W}^{} \nonumber \\
		&=& -\frac{1}{(4\pi)^2} \frac{g^4}{16 m_W^2} \left[\frac{10 c^2+1 }{c^2 } \Delta + \frac{18-24 c^2 }{s^2} \ln \left(\frac{\mu ^2}{m_W^2}\right) + \frac{14 c^4-9 c^2+1}{c^2 s^2} \ln\left(\frac{\mu ^2}{m_Z^2}\right) \right. \nonumber \\
		&& \left. + \frac{6}{s^2} \ln c^2 - \frac{1}{2 c^2} + 11 \right] - \frac{g^2}{4m_W^2} \left[\delta Z_e^{} - \frac{\delta s}{s} + \frac{1}{2} \delta Z_W^{} + \frac{1}{2} \left(\delta Z_{\nu_e^{}}^{\rm L} + \delta Z_{e}^{\rm L}\right)\right] \;, \nonumber \\
		C_{\nu^{}_e e W}^{} &=& 0 \;.
	\end{eqnarray}
	For CC interactions, the corrections to $A$ and $B$ are equal, whereas those to $C$ are vanishing. Meanwhile, from Fig.~\ref{fig:CC}, it is clear that there are two identical vertex corrections. Therefore, in the total amplitudes, the above expressions should be multiplied by a factor of two as we have already mentioned in connection with Eq.~(\ref{eq:A-CC}).
	
	\subsection{Box Diagrams}
	
	The last part of the one-loop amplitude is the contribution from box diagrams, which are actually UV-finite. The corresponding Feynman diagrams are drawn in Figs.~\underline{11} and \underline{12} for the NC and CC processes, respectively. The corrections to the coefficients from the NC process read
	\begin{eqnarray}
		A_{\rm \Box,NC}^{} &=& \frac{1}{(4\pi)^2} \frac{g^4}{2 m_W^2}\left[\frac{28 c^2-9}{16 c^2}+x_\alpha^{}\left(\ln x_\alpha^{}+1\right)\right] \;, \nonumber \\
		B_{\rm \Box,NC}^{} &=& \frac{1}{(4\pi)^2} \frac{g^4 }{2 m_W^2} \left[ \frac{ 24 c^4-20 c^2+15 }{16c^2} + x_\alpha^{} \left( \ln x_\alpha^{} + 1\right)\right] \;.
	\end{eqnarray}
	For the CC process, similar to the case of the $e$-$e$-$Z$ vertex, the corrections can also be divided into the EW and the QED part. The latter one comes from the box diagram which involves an internal photon propagator [see Fig.~\underline{12-(12)}]. The coefficients for the EW corrections are
	\begin{eqnarray}
		A^{\rm EW}_{\rm \Box,CC} = B^{\rm EW}_{\rm \Box,CC} = \frac{1}{(4\pi)^2} \frac{ g^4 \left(2 c^4+6 c^2-3\right)}{8 m_W^2 s^2}\ln \left(\frac{m_W^2}{m_Z^2}\right) \;,
	\end{eqnarray}
	while the corresponding terms from QED corrections are
	\begin{eqnarray}
		A^{\rm QED}_{\rm \Box,CC} &=& -\frac{1}{(4\pi)^2} \frac{g^4 s^2}{8 m_W^2 |{\bf p}_e^{}|} \left\{2 \ln \left(\frac{E_e^{}+|{\bf p}_e^{}|}{m_e^{}}\right) \left[3 E_e^{}+3 m_e^{}-2 E_e^{} \ln \left(\frac{m_e^{} |{\bf p}_e^{}|}{\lambda ^2}\right)\right] \right. \nonumber \\
		&& \left. -4 E_e^{} \ln \left(\frac{2 m_e^{}}{T_e^{}}\right) \ln \left(\frac{T_e^{}+|{\bf p}_e^{}|}{\sqrt{2 m_e^{} T_e^{}}}\right)+4 E_e^{} \Phi  +2 |{\bf p}_e^{}| \ln \left(\frac{m_W^2}{m_e^2}\right)+|{\bf p}_e^{}|\right\} \;, \nonumber \\
		B^{\rm QED}_{\rm \Box,CC} &=&  -\frac{1}{(4\pi)^2} \frac{g^4 s^2}{8 m_W^2 |{\bf p}_e^{}|} \left\{2 \ln \left(\frac{E_e^{}+|{\bf p}_e^{}|}{m_e^{}}\right) \left[3 E_e^{} -  m_e^{}-2 E_e^{} \ln \left(\frac{m_e^{} |{\bf p}_e^{}|}{\lambda ^2}\right)\right] \right. \nonumber \\
		&& \left. -4 E_e^{} \ln \left(\frac{2 m_e^{}}{T_e^{}}\right) \ln \left(\frac{T_e^{}+|{\bf p}_e^{}|}{\sqrt{2 m_e^{} T_e^{}}}\right)+4 E_e^{} \Phi +2 |{\bf p}_e^{}| \ln \left(\frac{m_W^2}{m_e^2}\right)+|{\bf p}_e^{}|\right\}  \;, \nonumber \\
		C^{\rm QED}_{\rm \Box,CC} &=& \frac{1}{(4\pi)^2} \frac{g^4 m_e^{} s^2 }{m_W^2 |{\bf p}_e^{}|} \ln \left(\frac{E_e^{}+|{\bf p}_e^{}|}{m_e^{}}\right)     \;,
	\end{eqnarray}
	with the function $\Phi$ introduced in Eq.~(\ref{eq:Phi}). 
	
	\subsection{Infrared Divergence}

	Thus far we have obtained the UV-finite corrections to the scattering amplitude. However, the QED corrections to the $e$-$e$-$Z$ vertex and the box diagram in the CC process suffer from the IR divergences when the fictitious mass of photon is set to zero, i.e., $\lambda \to 0$. These divergences are expected to be canceled {\it at the cross-section level} by taking account of the bremsstrahlung processes with real photon emission. 
	
	As is well known, the IR divergence in the NLO cross section due to the massless photon can be canceled by including the LO cross section for the processes where a soft photon is emitted from the final-state electron, i.e., when the photon energy lies below a certain energy cutoff. Hence the IR-finite cross section actually depends on the energy cutoff that varies for different experimental setups. If we further include the contributions from the hard photon emission, such a cutoff dependence disappears and the mass singularity or the collinear divergence in the limit of $m^{}_e \to 0$ is also removed. In this work, the momentum of the real photon will be integrated over its entire phase space such that both soft- and hard-photon emissions are taken into account. 
	
	\begin{figure}[t]
		\centering
		\includegraphics[scale=1]{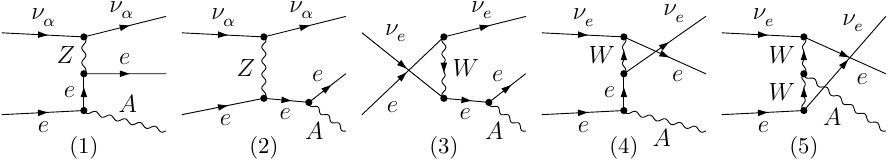}
		\caption{The Feynman diagrams of emitting a real photon in the NC process (1)-(2) and the CC process (3)-(5).}
		\label{fig:IR}
	\end{figure}
	
	The Feynman diagrams for real photon emissions from the initial- or final-state electrons and the intermediate $W$-boson are given in Fig.~\ref{fig:IR}. Only Fig.~\ref{fig:IR}-(1) and (2) contribute to the NC process for $\nu_{\mu,\tau}^{}$-$e$ scattering, while there are three extra diagrams from the CC interactions for $\nu_e^{}$-$e$ scattering. Compared to other processes that involve only one gauge-boson propagator, the amplitude of Fig.~\ref{fig:IR}-(5) is highly suppressed by the presence of two $W$-boson propagators which have no contribution to the IR divergence. Therefore, we do not take it into consideration.
	
	For the treatment of IR divergence, we adopt the results in Appendix B of Ref.~\cite{Sarantakos:1982bp} and generalize them to be applicable not only in the relativistic regime of the final-state electron, but also in the low-energy regime. However, it is worth noticing that the contribution of the photon bremsstrahlung to the total cross section is remarkable only when the energy of the final-state electron is high enough. Therefore, the extreme relativistic limit taken in Ref.~\cite{Sarantakos:1982bp} does not have a significant impact on the final numerical results. 
	
	The differential cross section for the $\nu_\mu^{}$-$e$ scattering with real photon emission at the tree level can be written as~\cite{Sarantakos:1982bp}	
	\begin{eqnarray}
		\frac{{\rm d} \sigma_{\gamma}^{(\mu)}}{{\rm d} T_e^{}} = \frac{g^6 s^2 m_e^{}}{256 \pi^3 m_W^4} \left\{\hat{R} \left[\frac{m_e^{} z }{E_\nu^{}}\left(c_{\rm A}^2-c_{\rm V}^2\right) + (c_{\rm A}^{} + c^{}_{\rm V})^2\right]+R (c_{\rm A}^{} - c^{}_{\rm V})^2 \right\} \;,
	\end{eqnarray}
	where the functions $R$ and $\hat{R}$ read
	\begin{eqnarray}
		R &=& (1-z)^2 \left\{ \frac{E_e^{}}{|{\bf p}^{}_e|}\ln \left(\frac{E_e^{}+|{\bf p}^{}_e|}{m_e^{}}\right) \ln \left(\frac{m_e^2}{\lambda^2}\right) -\ln \left(\frac{4 E_\nu^2}{\lambda^2}\right) + \frac{1}{2} \ln^2\left(\frac{E_e^{}+|{\bf p}^{}_e|}{m^{}_e}\right) \right. \nonumber \\
		&& \left. +\ln (1-z) \ln \left(\frac{E_e^{} + |{\bf p}^{}_e|}{m_e}\right) -\ln \left[z^2 (1-z)\right] -\frac{1}{2} \ln ^2(1-z) - {\rm Sp}\left(\frac{2|{\bf p}^{}_e|}{E_e^{} + |{\bf p}^{}_e|}\right) + 2\right\} \nonumber \\
		&& - z^2 \ln z \ln \left(\frac{E_e^{} + |{\bf p}^{}_e|}{m^{}_e}\right) + 3 z^2 \ln z + z^2 \left[{\rm Sp}\left(\frac{2 |{\bf p}^{}_e|}{E_e^{}+|{\bf p}_e^{}|}\right)-{\rm Sp}(z)\right] \nonumber \\
		&& -z(1-z) \ln \left(\frac{2 E_\nu^{}}{m_e^{}}\right) +2 z (1-z) + z(1-z) \ln (1-z) \;, \\
		\hat{R} &=& z \left[2 \ln \left(\frac{E_e^{} + |{\bf p}^{}_e|}{m_e^{}}\right) - \ln\left(\frac{2 E_\nu^{}}{m_e^{}}\right) -\ln (1-z)-2\right] + \frac{1}{2} \ln^2\left(\frac{E_e^{}+|{\bf p}^{}_e|}{m^{}_e}\right) \nonumber \\
		&& +\ln (1-z) \ln \left(\frac{E_e^{} + |{\bf p}^{}_e|}{m_e^{}}\right) + \frac{E_e^{}}{|{\bf p}^{}_e|}\ln \left(\frac{E_e^{} + |{\bf p}^{}_e|}{m_e^{}}\right)\ln \left(\frac{m_e^2}{\lambda^2}\right)-\ln \left(\frac{4 E_\nu^2}{\lambda^2}\right) \nonumber \\
		&& -\ln \left[z^2 (1-z)\right]-\frac{1}{2} \ln^2(1-z)- {\rm Sp}\left(\frac{2|{\bf p}^{}_e|}{E_e^{} + |{\bf p}^{}_e|}\right) +2 \;.
	\end{eqnarray}
	For the $\nu_e^{}$-$e$ scattering, one can simply take $c_{\rm V,A}^{} \to c_{\rm V,A}^{} + 1$ as we have already explained in Sec.~\ref{sec:strategy}. Similarly, for the antineutrino case, one should make the change of $c_{\rm A}^{} \to - c_{\rm A}^{}$. 
	
	After taking such contributions of real photon emissions into consideration, we arrive at the UV- and IR-finite cross sections. The total differential cross section at the one-loop level for the $\nu_\alpha^{}$-$e$ scattering is
	\begin{eqnarray}
		\frac{{\rm d} \sigma_{}^{(\alpha)}}{{\rm d} T_e^{}} = \frac{{\rm d} \sigma_{1}^{(\alpha)}}{{\rm d} T_e^{}} + \frac{{\rm d} \sigma_{\gamma}^{(\alpha)}}{{\rm d} T_e^{}} \;.
	\end{eqnarray}
	With the UV- and IR-finite corrections given in this and previous subsections, we now proceed to specify the input parameters and present the numerical values of the cross section.
	
	Finally, we make a brief comment on the gauge dependence of our results. If the calculations are performed in the general $R_\xi^{}$ gauge, each part of the contributions involving gauge bosons (i.e., the gauge-boson self-energies, the vertex corrections and the box diagrams) is gauge-dependent. As mentioned before, the 't Hooft-Feynman gauge with $\xi = 1$ has been chosen in our calculations. There exist extensive discussions about practically useful methods ensuring the gauge-independence of the results at the level of Green's functions (see, e.g., Refs.~\cite{Degrassi:1989ip,Degrassi:1992ff,Degrassi:1992ue}). However, at the level of cross sections, the final results are independent of the choices of gauge parameters.
	
	\section{Results and Applications}
	\label{sec:num}
	
	\subsection{Numerical Results}
	
	The latest values of input parameters are taken from Particle Data Group~\cite{ParticleDataGroup:2024cfk}, which are also summarized below
	\begin{itemize}
		\item The fine-structure constant
		\begin{eqnarray}
			\label{eq:alpha}
			\alpha \equiv e^2/(4\pi) = 1/137.035999084 \; ;
		\end{eqnarray}
		
		\item The gauge-boson and Higgs-boson masses\footnote{The central values of the $W$-boson mass from latest measurements in both ATLAS ($m_W^{\rm ATLAS} = 80.367~{\rm GeV}$)~\cite{ATLAS:2024erm} and CMS experiments ($m_W^{\rm CMS} = 80.360~{\rm GeV}$)~\cite{CMS:2024lrd} are consistent with the SM prediction. Therefore, we do not consider the $7\sigma$ discrepancy with the result from the CDF II collaboration~\cite{CDF:2022hxs}.}
		\begin{eqnarray}
			\label{eq:mwmzmh}
			m_W^{} = 80.369~{\rm GeV}\;, \quad \ m_Z^{} = 91.188~{\rm GeV} \;, \quad \ m_h^{} = 125.20~{\rm GeV}\;;
		\end{eqnarray}
		
		\item The quark masses
		\begin{eqnarray}
			&& m_u^{} = 62~{\rm MeV} \;, \quad m_c^{} = 1.67~{\rm GeV} \;, \quad m_t^{} = 172.57~{\rm GeV} \;, \nonumber\\
			&& m_d^{} = 83~{\rm MeV} \;, \quad m_s^{} = 215~{\rm MeV} \;, \quad m^{}_b = 4.78~{\rm GeV} \;;
		\end{eqnarray}
		
		\item The charged-lepton masses
		\begin{eqnarray}
			m_e^{} = 0.511~{\rm MeV} \;, \quad m_\mu^{} = 105.658~{\rm MeV} \;, \quad m_\tau^{}= 1.777~{\rm GeV} \;.
		\end{eqnarray}
	\end{itemize}
	For heavy quarks (i.e., $c$, $b$ and $t$) and charged fermions, the values listed above are actually the pole masses. However, we treat them as the on-shell masses in our one-loop calculations, since the difference between the pole mass and the on-shell mass appears at the two-loop level~\cite{Sirlin:1991fd,Sirlin:1991rt,Pilaftsis:1997dr}. For three light quarks (i.e., $u$, $d$ and $s$), due to the non-perturbative nature of strong interactions at low energies, they do not appear as real degrees of freedom and there will be no reliable on-shell masses to be extracted from measurements. Therefore, in this work, we adopt the effective quark masses from Refs.~\cite{Jegerlehner:1990uiq,Marciano:1993jd}, which have been implemented to account for the hadronic contributions to the vacuum polarization and evaluated from the measurements of $R \equiv \sigma\left(e^+ + e^- \to {\rm hadrons}\right) / \sigma\left(e^+ + e^- \to \mu^+ + \mu^-\right) $ via the dispersion relation. More detailed discussions can be found in Refs.~\cite{Marciano:1983ss,Hollik:1988ii,Denner:1991kt}.
	
	\begin{figure}[t]
		\centering
		\includegraphics[scale=0.87]{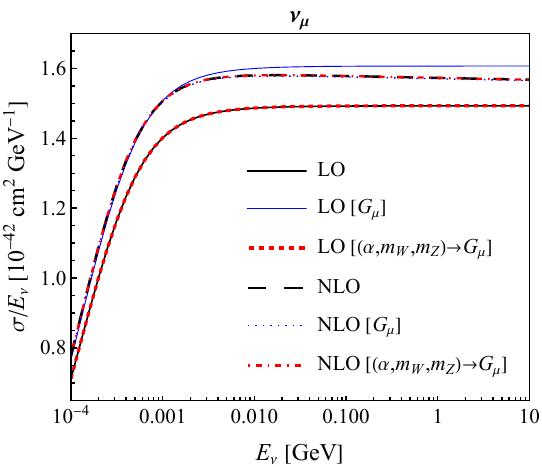}
		\includegraphics[scale=0.87]{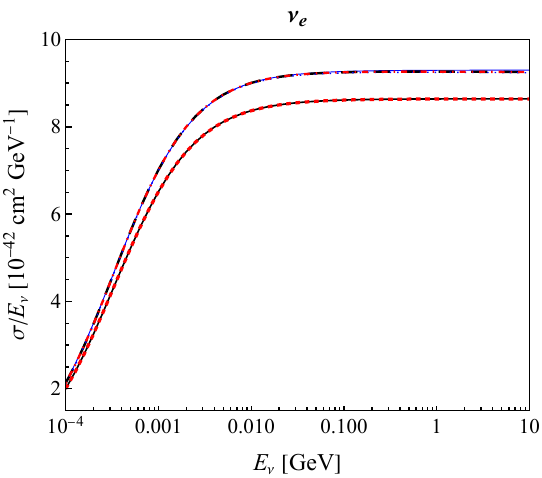}\\
		\includegraphics[scale=0.87]{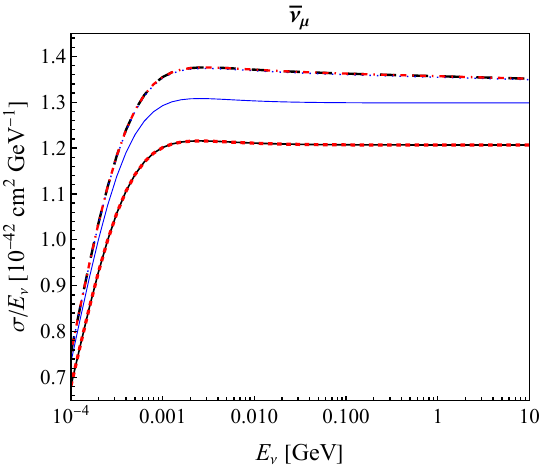}
		\includegraphics[scale=0.87]{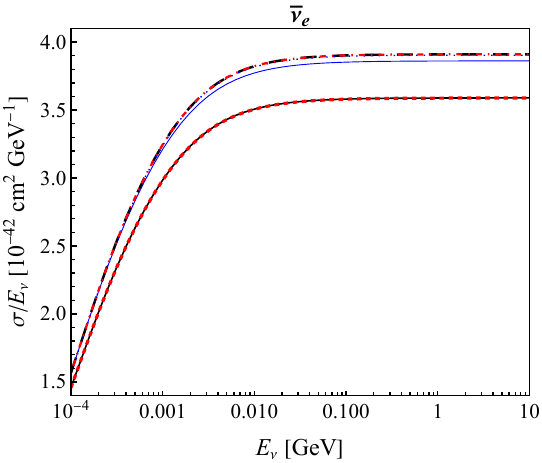}
		\vspace{-0.2cm}
		\caption{The total cross section $\sigma/E_\nu^{}$ as the function of the neutrino energy $E_\nu^{}$ at the LO (black solid curves) and the NLO (black dashed curves). The results in Ref.~\cite{Sarantakos:1982bp} are represented by blue and red curves, respectively. The blue ones are obtained by using $G^{}_\mu$ extracted from muon decays as the input parameter (solid curves for LO and dotted curves for NLO), the latter ones using $\alpha$, $m_W^{}$ and $m_Z^{}$ to calculate $G^{}_\mu$ (short-dashed curves for LO and dot-dashed curves for NLO).}
		\label{fig:our_MS}
	\end{figure}
	
	\begin{figure}[t]
		\centering
		\includegraphics[scale=0.87]{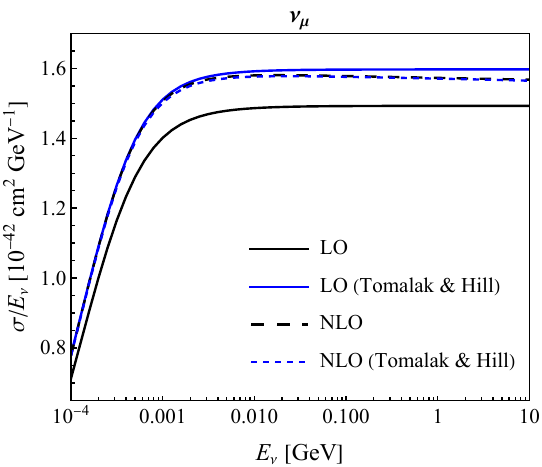}
		\includegraphics[scale=0.87]{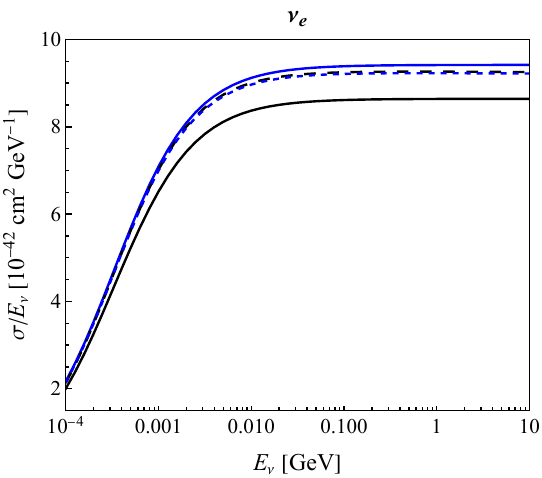}\\
		\includegraphics[scale=0.87]{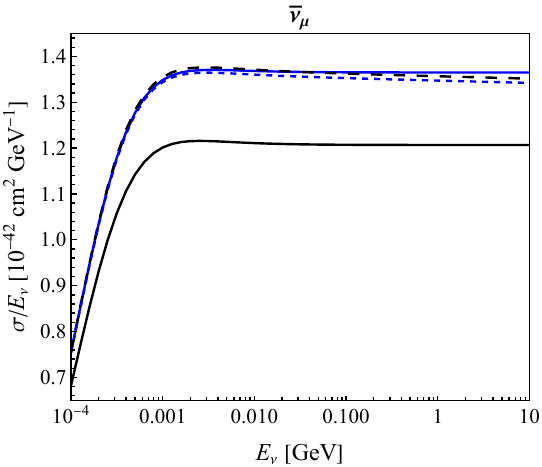}
		\includegraphics[scale=0.87]{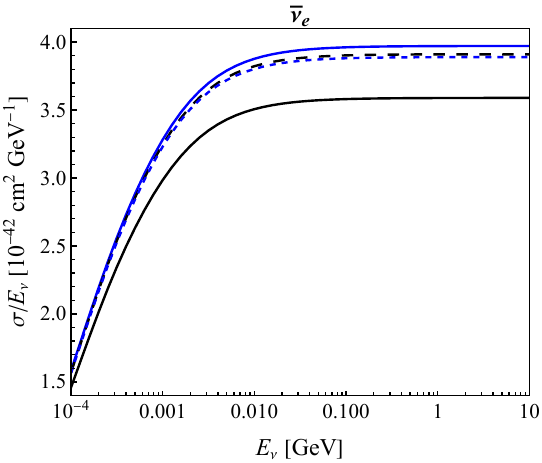}
		\vspace{-0.2cm}
		\caption{The total cross section $\sigma/E_\nu^{}$ in this work at the LO (black solid curves) and NLO (black dashed curves). As a comparison, the cross sections obtained from Ref.~\cite{Tomalak:2019ibg} are also plotted in blue solid and blue dashed curves for the LO and NLO results, respectively.}
		\label{fig:our_To}
	\end{figure}
	
	By incorporating all contributions to the coefficients $A$, $B$, and $C$ into Eq.~(\ref{eq:diff_Te_1}) and adding the contributions from the real photon emissions, we can integrate over the kinetic energy $T_e^{}$ to obtain the total cross section at the one-loop level. In Figs.~\ref{fig:our_MS} and \ref{fig:our_To}, we plot the numerical results of the total cross section for elastic (anti)neutrino-electron scattering $\sigma/E_\nu^{}$ as the function of the initial neutrino energy $E_\nu^{}$ at the LO (black solid curves) and the NLO (black dashed curves). For comparison, we also plot the numerical values of the cross sections from Ref.~\cite{Sarantakos:1982bp} in Fig.~\ref{fig:our_MS} and from Ref.~\cite{Tomalak:2019ibg} in Fig.~\ref{fig:our_To}, respectively. Some discussions about such a comparison are helpful.
	\begin{itemize}
		\item The cross section over neutrino energy $\sigma/E_\nu^{}$ approaches a constant when the initial neutrino energy is high enough, while it decreases with $E_\nu^{}$ at lower neutrino energies. This can be understood analytically. We take the tree-level cross section as an example, which is given in Eq.~(\ref{eq:diff_Te_0}). The explicit form of the total cross section reads
		\begin{eqnarray}
			\frac{\sigma_0^{(\mu)}}{E_\nu^{}} = \frac{g^4 m_e^{}}{64\pi m_W^4} \left[\frac{m_e^{} z_{\rm m}^2}{2 E_\nu^{}} \left(c_{\rm A}^2 - c^2_{\rm V} \right) + \frac{z_{\rm m}^{} }{3} \left(z_{\rm m}^2 - 3 z_{\rm m}^{} + 3 \right) (c_{\rm A}^{} - c^{}_{\rm V})^2 + z_{\rm m}^{} (c_{\rm A}^{} + c^{}_{\rm V})^2\right] \;. \quad
		\end{eqnarray}
		If the neutrino energy gets larger, we have $z_{\rm m}^{} = 1/\left[m_e^{}/(2 E_\nu^{}) + 1\right] \to 1$, then the first term in the square brackets, which is inversely proportional to $E_\nu^{}$, gradually approaches zero. Meanwhile, the last two terms remain constant, leading to a constant cross section at higher neutrino energies. As $E_\nu^{}$ gradually decreases, $z_{\rm m}^{} \to 2 E_{\nu}^{} / m^{}_e$ becomes vanishingly small. At this time, all three terms in the square brackets are proportional to $z_{\rm m}^{}$ such that the total cross section becomes smaller at lower neutrino energies. The turning point is at $E_\nu^{} \simeq m^{}_e \simeq {\cal O}({\rm MeV})$, which determines the relative size of those two terms in the denominator of $z^{}_{\rm m}$.
		
		\item In Fig.~\ref{fig:our_MS}, we present two sets of numerical results from Ref.~\cite{Sarantakos:1982bp}, represented by blue and red curves, respectively. The only difference between them lies in the choice of input parameters. For the results represented by blue curves, instead of the $W$-boson mass, we choose the Fermi constant $G_\mu^{} \approx 1.166 \times 10^{-5}~{\rm GeV}^{-2}$~\cite{ParticleDataGroup:2024cfk} from the muon lifetime as the input parameter at both LO and NLO. For the red curves, we use our input parameters $\alpha$, $m_W^{}$ and $m_Z^{}$ to calculate the corresponding $G_\mu^{}$ at both the tree and one-loop level. At the tree level, we have the relation
		\begin{eqnarray}
			\sqrt{2} G_\mu^{\rm LO} = \frac{\pi \alpha m_Z^2}{m_W^2 \left(m_Z^2 - m_W^2\right)} \;.
		\end{eqnarray}
		With the numerical values of $\alpha$, $m_W^{}$ and $m_Z^{}$ in Eqs.~(\ref{eq:alpha}) and (\ref{eq:mwmzmh}), we find the LO Fermi constant as $G_\mu^{\rm LO} \approx 1.124 \times 10^{-5}~{\rm GeV}^{-2}$. At the one-loop level, the relation is modified to
		\begin{eqnarray}
			\sqrt{2} G_\mu^{\rm NLO} = \frac{\pi \alpha m_Z^2}{m_W^2 \left(m_Z^2 - m_W^2\right)} \left(1+\Delta r \right)  \;.
		\end{eqnarray}
		With the transverse self-energies of gauge bosons $\Sigma_{\rm T}^A (q^2)$, $\Sigma_{\rm T}^W (q^2)$, $\Sigma_{\rm T}^Z (q^2)$ and $\Sigma_{\rm T}^{AZ}(q^2)$, the finite correction $\Delta r$ can be expressed as~\cite{Sirlin:1980nh,Denner:1991kt,Denner:2019vbn}
		\begin{eqnarray}
			\Delta r  &=& - \left. \frac{\partial \Sigma_{\rm T}^{A}\left(q^{2}\right)}{\partial q^{2}}\right|_{q^{2}=0} + \frac{c^2}{s^2}\left[\frac{\Sigma_{\rm T}^{Z} \left(m_Z^2\right)}{m_Z^2} - \frac{\Sigma_{\rm T}^{W} \left(m_W^2\right)}{m_W^2}\right] + \frac{\Sigma_{\rm T}^W(m_W^2) - \Sigma_{\rm T}^W(0)}{m_W^2} \nonumber \\
			&& - 2\frac{c}{s} \frac{\Sigma_{\rm T}^{AZ} (0)}{m_Z^2} + \frac{\alpha}{4\pi s^2}\left[6+\frac{7-4s^2}{2s^2}\ln\left(\frac{m_W^2}{m_Z^2}\right)\right] \;.
		\end{eqnarray}
		One can obtain that $\Delta r \approx 3.8\%$ and $G_\mu^{\rm NLO}\approx 1.167 \times 10^{-5}~{\rm GeV}^{-2}$, which is close to the value extracted from the muon lifetime. It can be observed that the tree-level results calculated by directly inputting $G_\mu^{}$ (blue solid curves) differ from ours, while the tree-level ones using the corrected $G_\mu^{}$ (red short-dashed curves) match ours very well. At the one-loop level, all three results are consistent with each other as the values of $G_\mu^{}$ in different cases also perfectly agree. 
		
		\item In Fig.~\ref{fig:our_To} we plot our numerical results together with those from Ref.~\cite{Tomalak:2019ibg}, which are calculated in the framework of the low-energy effective theory of the SM. At the tree level, there remains a notable difference between these two results. Similar to the case in Fig.~\ref{fig:our_MS}, it comes from the different choices of input parameters. At the one-loop level, the results turn out to be very close to each other, while small differences can be ascribed to different frameworks and input parameters. A more detailed comparison seems to be quite nontrivial, since one should take care of the conversion between on-shell and $\overline{\rm MS}$ parameters and renormalization-group running of the latter at the same order of perturbations. If all calculations were limited to the one-loop level, the results would be consistent. In reality, the difference arises from higher-order contributions in QCD, NLO logarithmic corrections, and small difference in modeling of the light-quark contributions.
	\end{itemize}
	
	It is worthwhile to stress that the determination of the relative size of the one-loop correction, i.e., whether the one-loop result is larger or smaller than the tree-level one, strongly depends on the theoretical framework and the choice of input parameters. For instance, let us compare the cross sections of $\nu_\mu^{}$-$e$ scattering in Fig.~\ref{fig:our_MS}. Our calculation shows that the tree-level result (black solid curve) is smaller than the one-loop level one (black dashed curve), indicating that the radiative correction is positive. On the other hand, if $G_\mu^{}$ instead of $m^{}_W$ is chosen as an input parameter, then a larger tree-level result (blue solid curve) is obtained, implying a negative one-loop correction. Similar circumstances also arise in Fig.~\ref{fig:our_To}, where not only is the value of $G_\mu^{}$ different from ours, but also is that of $\sin^2\theta_{\rm w}^{}$, making the discrepancy more pronounced. Admittedly, using $G_\mu^{}$ from the muon decay as an input parameter yields a relatively smaller correction and a more precise result compared to that with $m_W^{}$. However, without specifying the theoretical framework, naively stating the relative size and the sign of one-loop corrections is not physically meaningful. The main purpose of calculating higher-order corrections is to approach the true values of observables as closely as possible.
	
	\begin{figure}[t]
		\centering
		\includegraphics[scale=1]{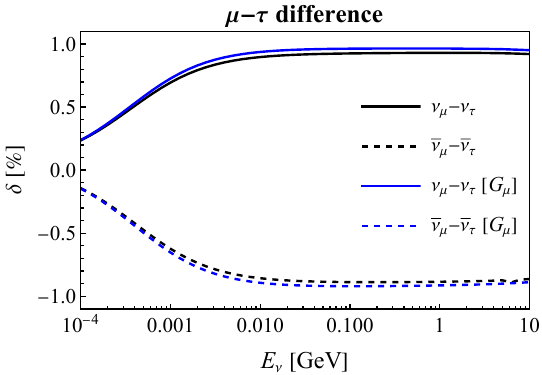}
		\vspace{-0.2cm}
		\caption{The differences in the cross sections of muon and tau neutrinos (black solid curve) and antineutrinos (black dashed curve). The results from Ref.~\cite{Sarantakos:1982bp} are plotted as blue curves.}
		\label{fig:mu_tau}
	\end{figure}
	
	In the above discussions, we focus on neutrinos and antineutrinos of electron and muon flavors. For completeness, we now make a brief comment on the differences in the NLO cross sections of muon and tau (anti)neutrinos. As the LO cross sections are equal for muon and tau flavors, namely, $\sigma^{(\mu)}_0 = \sigma^{(\tau)}_0$ and $\sigma^{(\overline{\mu})}_0 = \sigma^{(\overline{\tau})}_0$, one can define
	\begin{eqnarray}
		\label{eq:mutau}
		\delta^{}_{\mu-\tau} \equiv \left[\sigma^{(\mu)}_1 - \sigma^{(\tau)}_1\right]/\sigma^{(\mu)}_0 \; , \quad \delta^{}_{\overline{\mu}-\overline{\tau}} \equiv \left[\sigma^{(\overline{\mu})}_1 - \sigma^{(\overline{\tau})}_1\right]/\sigma^{(\overline{\mu})}_0 \; ,
	\end{eqnarray}
	to characterize the differences at the NLO. The numerical results of $\delta^{}_{\mu-\tau}$ and $\delta^{}_{\overline{\mu}-\overline{\tau}}$ are shown as black curves in Fig.~\ref{fig:mu_tau}, while the results from Ref.~\cite{Sarantakos:1982bp} are also plotted as blue curves for reference. Since the relative differences are independent of the value of $G_\mu^{}$, the discrepancy between black and blue curves is not induced by the input parameters. However, the discrepancy is at most of ${\cal O}(0.1\%)$. The solid curves indicate the difference in neutrinos, while the dashed ones correspond to that for antineutrinos. For $\nu_\mu^{}$ and $\nu_\tau^{}$ involving only NC interactions, the flavor-dependent difference at the NLO is of ${\cal O}(1\%)$, which is consistent with previous conclusions~\cite{Brdar:2023ttb}. The sign difference between the cross sections of neutrinos and antineutrinos can be understood as follows. The dominant contribution to the flavor-dependent corrections comes from the $\nu_\alpha^{}$-$\nu_\alpha^{}$-$A$ vertex diagrams. For the sake of clarity, we assume $m_e^{} \ll E_\nu^{}$ and replace the $q^2$-dependence in $A^{}_{\nu_\alpha^{} \nu_\alpha^{} A }$ by its average value $\overline{q}^2$~\cite{Sarantakos:1982bp,Marciano:2003eq}. Therefore, one may directly obtain the differences between muon and tau neutrinos by integrating Eq.~(\ref{eq:diff_Te_1}):
	\begin{eqnarray}
		\sigma_1^{(\mu)} - \sigma_1^{(\tau)} \simeq -  \frac{g^2 m_e^{} E_\nu^{}}{8\pi m_W^2} \times \frac{2}{3} \left( c_{\rm A}^{} + 2 c_{\rm V}^{} \right) \left[A^{}_{\nu_\mu^{} \nu_\mu^{} A } \left(\overline{q}^2 \right) - A^{}_{\nu_\tau^{} \nu_\tau^{} A } \left(\overline{q}^2 \right)\right] \;.
	\end{eqnarray}
	For antineutrinos, the result can be obtained via the replacement $c_{\rm A}^{} \to  - c_{\rm A}^{}$ as usual. With the help of the corresponding expressions of $c_{\rm V}^{}$ and $c_{\rm A}^{}$, one finds
	\begin{eqnarray}
		\frac{\sigma_1^{(\mu)} - \sigma_1^{(\tau)}}{\sigma_1^{(\overline{\mu})} - \sigma_1^{(\overline{\tau})}} \simeq \frac{3-8s^2}{1-8s^2} <0 \;, 
	\end{eqnarray}
	where $s^2 \approx 0.223$ in the on-shell scheme should be noticed. Therefore, there is the sign difference between the cases of neutrinos and antineutrinos in Fig.~\ref{fig:mu_tau}.
	
	\subsection{Reactor Antineutrinos}
	As a practical application, considering one-loop corrections to the cross sections, we compute the event rates of the elastic $\overline{\nu}$-$e$ scattering in next-generation reactor neutrino experiments. The electron antineutrinos $\overline{\nu}^{}_e$ from nuclear reactors are created in the fission of four dominant isotopes, i.e., $^{235}{\rm U}$, $^{238}{\rm U}$, $^{239}{\rm Pu}$ and $^{241}{\rm Pu}$. The antineutrino flux $\phi_i^{} (E_\nu^{})$ from the reactor $i$ with the thermal power $P_i^{}$ can be calculated as
	\begin{eqnarray}\label{eq:flux}
		\phi_i^{} (E_\nu^{}) = \frac{P_i^{}}{\displaystyle \sum_j f_j^{} \epsilon_j^{}} \sum_j f_j^{} S_j^{}(E_\nu^{}) \;,
	\end{eqnarray}
	where $f_j^{}$, $\epsilon_j^{}$ and $S_j^{}(E_\nu^{})$ are the fission fraction, the thermal energy released in each fission, and the neutrino flux per fission for the $j$-th isotope, respectively. In this work, the values of the first two parameters are adopted from Refs.~\cite{JUNO:2015zny,JUNO:2020ijm}, as listed in Table~\ref{tab:isotope} for reference. The spectrum can be expressed with a 5th order polynomial parametrization~\cite{Mueller:2011nm}
	\begin{eqnarray}
		S_j^{}(E_\nu^{}) = \exp\left(\sum_{p=1}^{6} \alpha_{pj}^{} E_\nu^{p-1}\right) \;,
	\end{eqnarray}
	where the corresponding coefficients for the isotope $j$ at the $(p-1)$-th order $\alpha_{pj}^{}$ can be obtained from the Table~VI of Ref.~\cite{Mueller:2011nm}. 

	\begin{table}[t]
		\centering
		\begin{tabular}{ccccc}
			\hline\hline
			& $^{235}{\rm U}$ & $^{238}{\rm U}$ & $^{239}{\rm Pu}$ & $^{241}{\rm Pu}$ \\ \hline
			$f^{}_i$   &  0.561   &  0.076   &  0.307   &  0.056   \\
			$\epsilon^{}_i/{\rm MeV}$   &  202.36   &  205.99   &  211.12   &  214.26   \\ \hline\hline
		\end{tabular}
		\caption{The fission fraction $f^{}_i$ and the thermal energy $\epsilon^{}_i$ released in each fission for four main isotopes in nuclear reactors~\cite{JUNO:2015zny,JUNO:2020ijm}.}
		\label{tab:isotope}
	\end{table}	

	As the electron antineutrinos travel from the reactor to the detector, they may change from one flavor to another. The oscillation probability $P_{\alpha\beta}^{} (E,L) \equiv P(\overline{\nu}_\alpha^{} \to \overline{\nu}_\beta^{})$ of reactor antineutrinos with the energy $E$ and the distance $L$ between the nuclear power plants (NPPs) and the detector is given by
	\begin{eqnarray}
		\label{eq:oscillation_prob}
		P_{\alpha\beta}^{} (E,L) &=& \delta^{}_{\alpha \beta} - 4\sum_{i<j}^3{\rm Re}\left(U_{\alpha i}^{} U_{\beta j}^{} U_{\alpha j}^* U_{\beta i}^*\right) \sin^2\left(\frac{\Delta m_{ji}^2 L}{4E}\right) \nonumber \\
		&& - 8 {\cal J} \sum_\gamma \epsilon^{}_{\alpha\beta\gamma} \sin\left(\frac{\Delta m_{21}^2 L}{4E}\right) \sin\left(\frac{\Delta m_{31}^2 L}{4E}\right) \sin\left(\frac{\Delta m_{32}^2 L}{4E}\right) \;,
	\end{eqnarray}
	where $\Delta m_{ji}^2 \equiv m_j^2 - m_i^2$ (for $ji=21,31,32$) are neutrino mass-squared differences, $U_{\alpha i}^{}$ (for $\alpha = e,\mu,\tau$ and $i=1,2,3$) denotes the elements of leptonic flavor mixing matrix~\cite{Pontecorvo:1957qd,Maki:1962mu,Pontecorvo:1967fh}, and $\epsilon^{}_{\alpha\beta\gamma}$ is the three-dimensional Levi-Civita symbol. As advocated by Particle Data Group~\cite{ParticleDataGroup:2024cfk}, $U$ is parametrized by three mixing angles $\theta_{ij}^{}$ ($ij=12,13,23$) and one Dirac CP-violating phase $\delta_{\rm CP}^{}$. Two possible Majorana-type CP phases are irrelevant for ordinary neutrino oscillations. The Jarlskog invariant can be expressed as ${\cal J} = \sin\theta_{12}^{} \cos\theta_{12}^{} \sin\theta_{23}^{} \cos\theta_{23}^{} \sin\theta_{13}^{} \cos^2\theta_{13}^{} \sin\delta_{\rm CP}^{}$~\cite{Jarlskog:1985ht,Wu:1985ea}. 
	
	For the detector with the total electron number $N_e^{}$ and data-taking time $t$, the observed event spectrum of the elastic $\overline{\nu}_\alpha^{}$-$e$ scattering with the differential cross section ${\rm d}\sigma_{}^{(\overline{\alpha})} /{\rm d}T_e^{}$ reads~\cite{Li:2017dbg}
	\begin{eqnarray}
		\frac{{\rm d}N_\alpha^{}}{{\rm d}E_{\rm obs}^{}} = N_e^{} t \sum_i \int_{0}^{\infty} {\rm d}T_e^{}\ {\cal G}\left(E_{\rm obs};T_e^{},\delta^{}_{\rm E}\right) \int_{E_{\nu}^{\rm min}}^{\infty} P^{}_{e\alpha}(E_{\nu}^{},L_i^{}) \frac{\phi_i^{}(E_\nu^{})}{4\pi L_i^2} \frac{{\rm d}\sigma^{(\overline{\alpha})}(E_\nu^{})}{{\rm d}T_e^{}}\ {\rm d}E_\nu^{} \;,
	\end{eqnarray}
	where $E_{\rm obs}^{}$ stands for the observed energy, and the response function ${\cal G}\left(E_{\rm obs};T_e^{},\delta^{}_{\rm E}\right)$ is taken as the Gaussian function of $E_{\rm obs}^{}$ with the electron recoil energy $T_e^{}$ and the energy resolution of the detector $\delta_{\rm E}^{}$ being the expectation value and the standard deviation, respectively. The minimal neutrino energy $E_\nu^{\rm min} = T_e^{}/2 + \sqrt{T_e^{}\left(2m_e^{}+T_e^{}\right)}/2$ can be obtained from the kinematics for a given $T_e^{}$. Finally, one should sum up the contributions from all the NPPs with the distance $L_i^{}$ and the antineutrino flux $\phi_i^{}(E_\nu^{})$, while the flux and the oscillation probability can be read off from Eqs.~(\ref{eq:flux}) and (\ref{eq:oscillation_prob}). 
	
	Two reactor neutrino experiments will be considered. The first is Jiangmen Underground Neutrino Observatory (JUNO) experiment, which aims at determining neutrino mass ordering and precisely measuring oscillation parameters~\cite{JUNO:2015zny,JUNO:2022mxj}. Its central detector is located around 52.5 km away from both the Taishan (TS) and Yangjiang (YJ) NPPs, which accordingly has two 4.6 GW cores and six 2.9 GW cores. With a 20 kiloton liquid-scintillator (LS) target and a 12\% hydrogen mass fraction, the electron number is about $N_e^{} \approx 6.5 \times 10^{33}$. The energy resolution is expected to be better than 3\% at 1 MeV.\footnote{The latest analysis shows that the energy resolution can now reach 2.95\% at 1 MeV~\cite{JUNO:2024fdc}.}
	
	The second is Taishan Antineutrino Observatory (TAO), as a near detector of JUNO, is 44 m away from Taishan NPP reactor core-1 (TS-C1) and 217 m from the other one (TS-C2)~\cite{JUNO:2020ijm}. It is built to provide a reference antineutrino spectrum for the JUNO detector in order to eliminate the impact of possible fine structures that can mimic neutrino oscillation patterns. The central detector contains 2.8 tons of LS with the number of electrons $N_e^{} \approx 9.3 \times 10^{29}$. To study the fine structure of the antineutrino spectrum, the energy resolution is designed to be as high as 1.5\% at 1~MeV~\cite{JUNO:2020ijm}. 
	
	\begin{figure}[t]
		\centering
		\includegraphics[height=6.8cm]{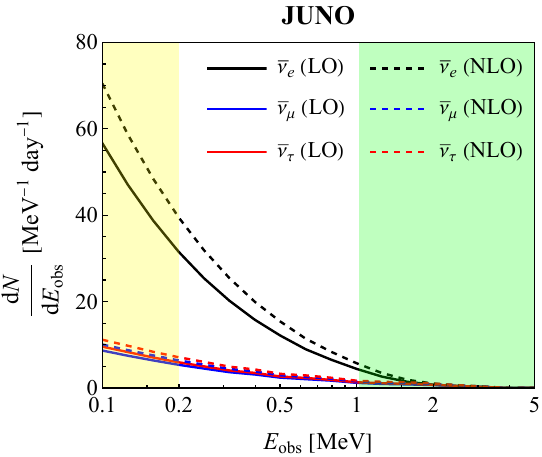}
		\includegraphics[height=6.8cm]{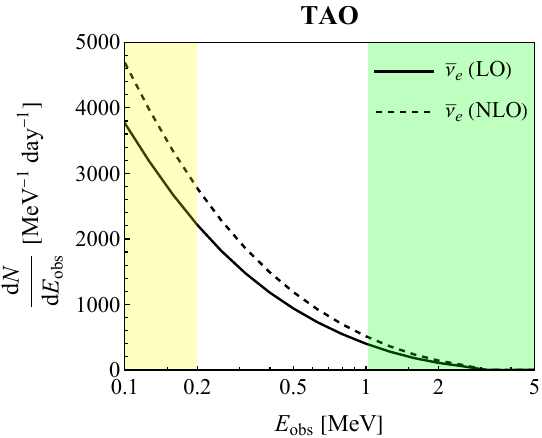}
		\vspace{-0.2cm}
		\caption{The observed event spectrum per day of the $\overline{\nu}^{}_\alpha$-$e$ ($\alpha=e,\mu,\tau$) scattering in JUNO (left panel) and $\overline{\nu}_e^{}$-$e$ scattering in TAO (right panel). The spectra of the three flavors of antineutrinos are separately plotted as black, blue and red curves. Solid curves represent the spectra calculated with the LO cross sections, while the dashed curves with the NLO ones. The yellow and the green shadows denote the regions of $E_{\rm obs}^{} \lesssim 0.2~{\rm MeV}$ and $E_{\rm obs}^{} \gtrsim 1~{\rm MeV}$, where the backgrounds may be dominant over signals.}
		\label{fig:events}
	\end{figure}
	
	With the above experimental setups, we are ready to numerically calculate the event rates of $\overline{\nu}^{}_\alpha$-$e$ scattering at both JUNO and TAO. For definiteness, we take the latest best-fit values of neutrino oscillation parameters in the case of normal mass ordering ($m_1^{} < m_2^{} < m_3^{}$) from Ref.~\cite{Esteban:2024eli}. The event spectra are plotted in Fig.~\ref{fig:events} for JUNO (left) and TAO (right), respectively. As a rough analysis of possible backgrounds for the signals of recoiled electrons in the LS detectors, we notice that for the observed energy $E_{\rm obs}^{} \lesssim 0.2~{\rm MeV}$ there will be huge radioactive backgrounds and dark noises, while for $E_{\rm obs}^{} \gtrsim 1~{\rm MeV}$ the misidentification of inverse-beta-decay (IBD) events $\overline{\nu}^{}_e + p \to e^+ + n$ serves as a dominant background~\cite{JUNO:2015zny}. Hence these two regions are denoted apart with yellow and green shadows in Fig.~\ref{fig:events}. Some comments on the event spectrum are in order.
	
	\begin{itemize}
		\item It is interesting to mention that $\overline{\nu}^{}_\mu$ and $\overline{\nu}^{}_\tau$ appear in the JUNO detector due to significant oscillations of reactor antineutrinos for the chosen baseline. The event spectra for $\overline{\nu}^{}_e$, $\overline{\nu}^{}_\mu$ and $\overline{\nu}^{}_\tau$ at JUNO are plotted in black, blue and red curves in the left panel of Fig.~\ref{fig:events}. The solid curves represent the spectra calculated with the LO cross sections, while the dashed ones with the NLO ones. In the range of $0.2~{\rm MeV} \lesssim E_{\rm obs}^{} \lesssim 1~{\rm MeV}$, there are approximately 12 events per day for $\overline{\nu}^{}_e$, while for $\overline{\nu}^{}_\mu$ and $\overline{\nu}_\tau^{}$ there are about 2 events per day. The difference between the $\overline{\nu}^{}_\mu$ and $\overline{\nu}_\tau^{}$ event spectra is negligibly small.
		
		\item Due to the short baseline, reactor antineutrinos do not oscillate significantly when arriving at TAO. Therefore, we only consider the $\overline{\nu}_e^{}$-$e$ scattering events. The event rate is about 900 per day, which is much larger than those at JUNO. Thanks to the high energy resolution and sufficiently large event rates, one can achieve precise measurements of the neutrino spectrum. On the other hand, the $\overline{\nu}^{}_e$ spectrum can be accurately measured in the IBD channel. With the one-loop cross sections as the input quantities for analyzing the experimental data, we can place more restrictive constraints on possible new-physics scenarios for neutrino interactions.
	\end{itemize}
	Before ending this section, we point out that the detection of reactor antineutrinos in JUNO and TAO is mainly based on the IBD events, and it is difficult to distinguish the final-state electrons from photons of low energies. Furthermore, it is very challenging to observe the recoiled electrons due to enormous radioactive backgrounds for single events. However, the detection of recoiled electrons from $\overline{\nu}^{}_\mu$ and $\overline{\nu}^{}_\tau$ will turn JUNO into an appearance experiment. In such a case, one may predict the $\overline{\nu}^{}_e$-$e$ background by first reconstructing the $\overline{\nu}^{}_e$ flux from the IBD events (at both JUNO and TAO) and then calculating the event rate of $\overline{\nu}^{}_e$-$e$ scattering. A dedicated study of the $\overline{\nu}^{}_\mu$ and $\overline{\nu}^{}_\tau$ appearance in JUNO and the experimental sensitivity of TAO to the cross section of $\overline{\nu}^{}_e$-$e$ scattering will be left for future works.

	\section{Summary}
	\label{sec:sum}
	
	In this paper, we have carried out an independent and complete calculation of the differential cross sections for elastic neutrino-electron scattering at the one-loop level in the SM. First, we accomplish the one-loop renormalization of the SM in the on-shell scheme and choose the fine-structure constant $\alpha$ and the on-shell masses $\{m^{}_W, m^{}_Z, m^{}_h, m^{}_f\}$ of the weak gauge bosons, the Higgs boson and the SM fermions as input parameters. By using the latest values of the input parameters, we numerically calculate the differential cross sections and make a comparison between our results and the previous ones in the literature. An excellent agreement has been found if the same scheme and input parameters are adopted. Our calculation will be useful in the coming era of precision neutrino physics when searching for the new-physics effects with the precision data. Second, as a practical application, we compute the event rates for the elastic scattering of reactor antineutrinos in the forthcoming JUNO and TAO experiments. We point out that the elastic neutrino-electron scattering can be implemented to probe the appearance of $\overline{\nu}^{}_\mu$ and $\overline{\nu}^{}_\tau$ in the JUNO detector. However, a realistic analysis of the relevant backgrounds in JUNO deserves further dedicated studies. For the TAO detector, as the event rate is large enough, it is promising to precisely measure the cross section of elastic neutrino-electron scattering. The combined analysis of both elastic neutrino-electron scattering and the IBD process will be helpful.

	Although we have concentrated on reactor neutrinos, the results obtained in this work are also applicable to the detection of supernova neutrinos and accelerator neutrinos in a broad range of neutrino energies. While JUNO is expected to improve the precision of oscillation parameters $\Delta m_{31}^2$, $\Delta m_{21}^2$ and $\theta_{12}^{}$ to the sub-percent level~\cite{JUNO:2022mxj}, the accelerator neutrino oscillation experiments DUNE~\cite{DUNE:2015lol} and T2HK~\cite{Hyper-Kamiokande:2018ofw} are aiming to determine the octant of $\theta^{}_{23}$ and measure the leptonic CP-violating phase $\delta_{\rm CP}^{}$ with a reasonably high accuracy. Theoretical calculations of the observables in the SM by including radiative corrections become indispensable. In particular, precise calculations in the neutrino sector will enable future neutrino experiments to draw more stringent constraints on the new-physics scenarios beyond the SM~\cite{Ellis:2020hus,Parke:2024xre,Herrera:2024ysj,Alves:2024twb,Brdar:2024lud,Giunti:2024gec}.

	\section*{Acknowledgments}
	We would like to thank the anonymous referee for many useful suggestions and comments which have helped us improve our manuscript. This work was supported in part by the National Natural Science Foundation of China under grant No.~12475113, and by the CAS Project for Young Scientists in Basic Research (YSBR-099). All Feynman diagrams in this work are generated by {\tt FeynArts}~\cite{Hahn:2000kx}, and loop integrals are calculated with the help of {\tt Package-X}~\cite{Patel:2015tea,Patel:2016fam}.

	\bibliographystyle{elsarticle-num}
	\bibliography{nu_e_scattering_ref}

\end{document}